\documentclass[acmsmall]{acmart}
\acmJournal{PACMHCI}

\usepackage{flushend}
\usepackage{float}
\usepackage{subfigure}
\usepackage{longtable}
\usepackage{booktabs}
\usepackage{multirow}
\AtBeginDocument{%
  \providecommand\BibTeX{{%
    \normalfont B\kern-0.5em{\scshape i\kern-0.25em b}\kern-0.8em\TeX}}}
\usepackage{tabularx}
\usepackage{bbding}
\usepackage{graphicx}
\usepackage{caption}
\usepackage{geometry}
\usepackage{amsmath}
\usepackage{makecell}
\usepackage{float}
\usepackage{color}
\usepackage{booktabs}
\usepackage[normalem]{ulem}
\usepackage{tabu}
\usepackage{hyperref}
\usepackage{xcolor}
\usepackage{appendix}
\usepackage{fancyvrb}
\usepackage{listings}

\setcopyright{acmlicensed}
\copyrightyear{2024}
\acmYear{2024}
\acmDOI{XXXXXXX.XXXXXXX}

\acmConference[Conference acronym 'XX]{Make sure to enter the correct
  conference title from your rights confirmation emai}{June 03--05,
  2018}{Woodstock, NY}
\acmISBN{978-1-4503-XXXX-X/18/06}

\newcommand{\removed}[1]{}           
\newcommand{\replace}[2]{#2}         
\newcommand{\added}[1]{#1}           

\author{Chu Zhang}
\authornote{Equal contribution.}
\email{zhangchu0908@outlook.com}
\orcid{0009-0004-4491-2279}
\affiliation{
\institution{City University of Hong Kong}
\city{Hong Kong}
\country{China}}

\author{XiaoKe Zeng}
\authornotemark[1]
\email{zxioke@outlook.com}
\orcid{0009-0007-5987-502X}
\affiliation{
\institution{City University of Hong Kong}
\city{Hong Kong}
\country{China}}

\author{Jin Zhang}
\email{stberries0@gmail.com}
\orcid{0009-0000-3849-4868}
\affiliation{
\institution{University of York}
\city{York}
\country{United Kingdom}}

\author{Ruoyu Wen}
\email{rwe77@uclive.ac.nz}
\orcid{0009-0008-0052-0045}
\affiliation{
\institution{University of Canterbury}
\city{Christchurch}
\country{New Zealand}}

\author{Vince Siu}
\email{vince@pressstartacademy.com}
\orcid{0009-0008-0391-6029}
\affiliation{
\institution{Press Start Academy}
\city{Hong Kong}
\country{China}}

\author{Richard William Allen}
\email{rwallen@cityu.edu.hk}
\orcid{0000-0003-2826-0990}
\affiliation{
\institution{City University of Hong Kong}
\city{Hong Kong}
\country{China}}

\author{RAY LC}
\authornote{Correspondences can be addressed to ray.lc@cityu.edu.hk.}
\email{ray.lc@cityu.edu.hk}
\orcid{0000-0001-7310-8790}
\affiliation{
\institution{City University of Hong Kong Studio for Narrative Spaces}
\city{Hong Kong, SAR}
\country{China}}

\begin{document}
\title[FIERO]{FIERO: Empowering Creative Writing Through Collaborative Game Play}

\begin{teaserfigure}
\centering
    \includegraphics[width=1\linewidth]{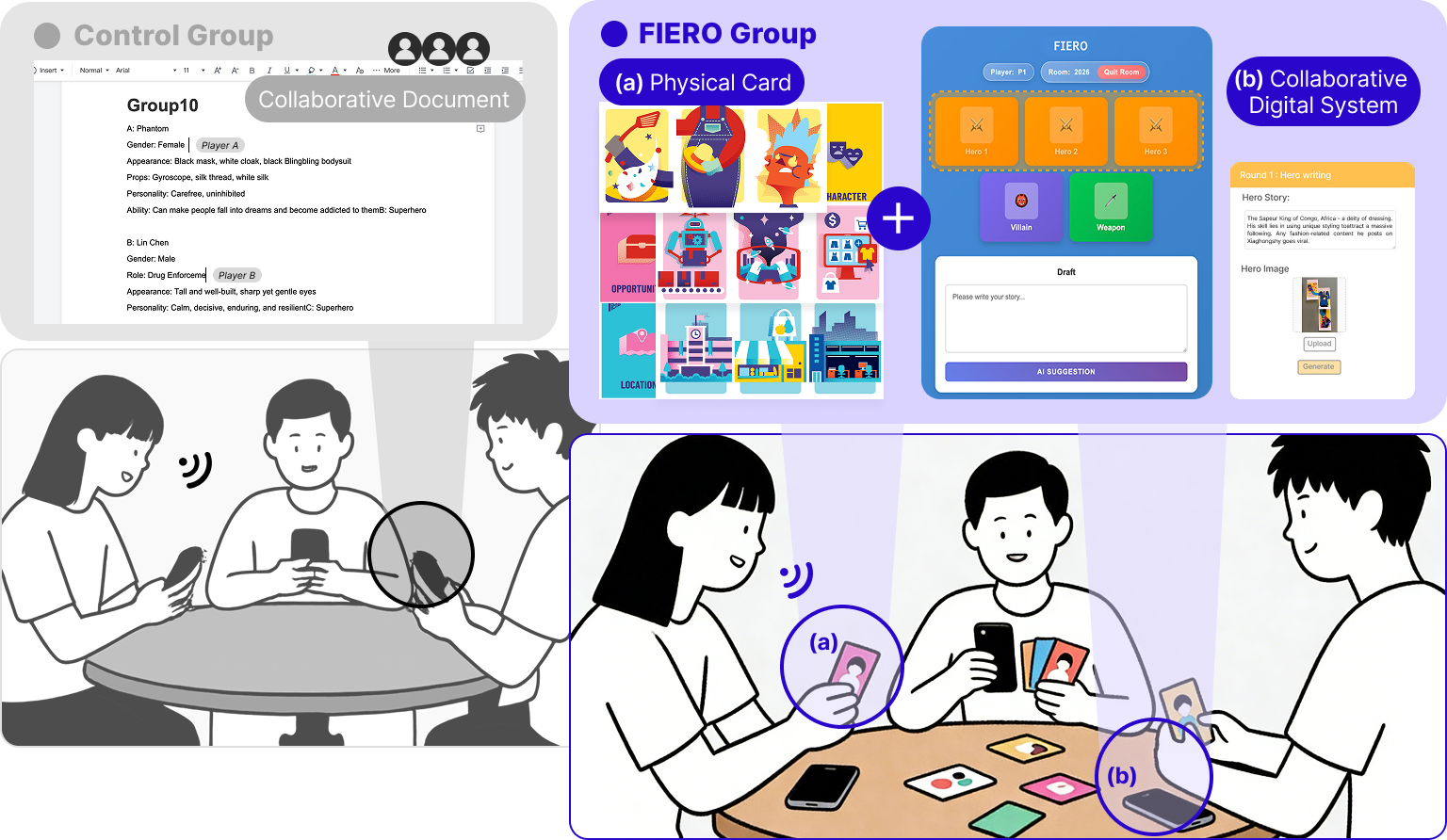}
    \vspace{-0.3cm}
    \caption{FIERO is a collaborative story creation game that combines card-based mechanics with digital system. Compared to traditional shared document-based collaboration (Control Group), FIERO helps participants develop a stronger sense of immersion and visualization in their story creation process, while enhancing the coherence, logic, and innovation of the story.}
    \label{fig:cover}
\end{teaserfigure}

\begin{abstract}
Creativity often flourishes in collaboration, such as when designers brainstorm a new app together, or storytellers collectively build a world with elements of each person’s narrative. However, collaborative storytelling can have challenges for its participants, such as when they disagree about the plot proposed, or when different ideas become fragmented when voiced individually. While current tools for creative collaboration focus on synchronous online text sharing, they often neglect the social dynamics of in-person collaboration critical to creative synergy. To address this, we created FIERO, a multiplayer web-based card game. Physical cards provide tangible scaffolding and social interaction, while the digital interface generates contextual visuals, facilitate group decisions, ensure narrative coherence, and synthesize different idea contributions using generative AI. Compared against online collaborative writing alone, the game significantly enhanced intuitive stimulation, idea fluency, and novelty generation, and also improved the content of the stories produced, leading to greater plot coherence (N=60). The cards provided creative structure and social engagement, while the interface provided contextualized augmentation without affecting player agency. This work shows how collaborative play can be utilized to foster creative support.

\end{abstract}




\maketitle

\section{Introduction}\label{sec:Introduction}
Creative writing is an activity of psychological value and cognitive benefit, requiring the generation of novel narratives, engaging characters, and coherent storylines \cite{chittooran2015reading}. However, individual writers often face creative blocks and limited perspectives that constrain their creative potential~\cite{sharples2002we}, leading them to abandon their ideas even before they begin \cite{karwowski2018measuring}. These challenges underscore the need for supportive creative environments \cite{sharples2002we}.

Existing research has sought to alleviate the writing burden by introducing moderate constraints (e.g., random vocabulary or visual prompts). Yet, these approaches often overlook the potential of collaboration in distributing the pressure of creation \cite{vass2008discourse}. In creative writing, collaboration not only breaks individual perspective limitations and stimulates divergent thinking \cite{lingard2021collaborative, sharples1996account} but also helps eliminate individual creative fears.

Existing collaborative tools (such as Google Docs) support synchronous editing, but they are built on the assumption of free editing, relying on users to spontaneously coordinate creative ideas \replace{without structured guidance}{with minimal structured guidance}. Meanwhile, in the absence of an effective integration mechanism, diverse perspectives may lead to conflicts, and the cognitive burden of reconciling different creative visions can undermine collaborative efficiency \cite{sharples1996account, lingard2021collaborative}. Consequently, a key challenge lies in synchronizing creative ideas within team collaboration to enhance the quality and experience of collaborative creation.

Tabletop games, particularly narrative-driven ones, serve as an ideal vehicle for creators by integrating rule-based constraints, playful exploration, and collective storytelling \cite{cover2014creation}. Meanwhile, the intervention of Artificial Intelligence (AI) provides technical possibilities for managing complex group negotiations \cite{qin2024charactermeet, chung2025toyteller, suh2024luminate}. \replace{To date, no}{To our knowledge, little} research has explored the synthesis of these two domains to support creative writing.

To address this gap, we present \textbf{FIERO}, \textit{a multiplayer web-based card game}. In this game, players co-construct narratives by combining physical cards representing heroes, villains, and weapons, and determine the overall plot trajectory through three rounds of negotiation and discussion. Throughout the process, teams can invoke generative AI to facilitate collaborative decision-making and enhance the narrative, further \replace{realizing the collective sense-making of the story}{enhancing the quality of collaborative writing}. Through the deployment and evaluation of FIERO, this paper seeks to address the following research questions:

\begin{itemize}
    \item RQ1: How do people engage in collaborative writing in the game play experience?
    \item RQ2: How do the stories created by collaboration in the game differ from those created by traditional collaborative writing?
\end{itemize}

In this paper, we present the design and evaluation of FIERO, demonstrating its effectiveness in enhancing creative writing experiences and story quality through a comparative experiment (N=60). Our findings show that this multi-sensory interaction transforms visual content into shared narrative anchors, significantly increasing intuitive stimulation during writing. Meanwhile, FIERO outperformed traditional collaborative writing in dimensions such as idea fluency and novelty generation, \replace{indicating its capacity to effectively empower users' creative writing processes}{suggesting that FIERO can empower users' creative writing processes in these aspects}. The improvement in player satisfaction was not only reflected across various dimensions of subjective questionnaires but also evidenced in content analysis, where the FIERO group showed significant advantages in plot coherence, character fidelity, and overall preference.

FIERO advances collaborative storytelling by demonstrating how a gamified framework can scaffold creative writing without compromising social interaction. Its structured gameplay combines card-based constraints, multisensory interaction, and iterative decision-making to convert abstract concepts into shared narrative anchors. This work provides valuable insights for future creativity support tools designed to strike a balance between guidance and creative freedom.

\section{Related Work}\label{sec:Background}
\subsection{Challenges in Creative Writing}

Creative writing has been linked to a range of psychological and cognitive benefits. It assists individuals in processing emotions, alleviating stress, and enhancing psychological resilience, serving as a valuable means for identity construction and the cultivation of reflective thinking \cite{chittooran2015reading}. It allows creators to explore different facets of the self within fictional realms. Furthermore, from the perspective of cognitive development, the creative process involves high-level divergent thinking and complex narrative construction, which significantly exercises the brain's associative capabilities and problem-solving skills \cite{runco2010divergent}.

However, many people struggle during the initial conceptualization phase. Faced with infinite creative possibilities and a lack of structured starting points, they often feel at a loss. A key factor hindering creative expression is the lack of creative self-efficacy \cite{karwowski2018measuring}, the belief that one lacks the ability to produce creative outcomes, which leads to evaluation apprehension. This psychological pressure causes users to pre-emptively negate their own ideas for fear of producing ``low-quality'' work, often leading to abandonment before the first word is even written \cite{osborn1963applied}. Such individual creative dilemmas highlight the limitations of solitary, unstructured creation modes.

To address creative dilemmas, existing research \replace{primarily}{often} introduces moderate constraints to reduce individual cognitive load. For instance, using random vocabulary, visual images, or specific story starters as creative prompts \cite{han2024AI, Gero2019, Yang2022AIAA} narrows the selection space to circumvent the logical pressure of ``the first line.'' With the evolution of human-AI co-creation technology, this constrained creation has shifted toward generative assistance. This model allows AI to pre-generate drafts or outlines \cite{Liu2025}; for creators, selecting or modifying existing material imposes far less psychological pressure than original creation from scratch. \added{Recent research demonstrates that incorporating multimodal elements such as sketches and images into story writing effectively supports divergent narrative exploration \cite{fu2026vistoria}. Card-based ideation tools are also widely applied in creative support due to their ideation-facilitating functions, such as providing structured conceptual frameworks \cite{rogerson2022smeft}, serving as convenient mediums for information transmission \cite{roy2019card}, and acting as cognitive bridges \cite{nurain2024designing}.} Beyond technical support, the introduction of a social dimension is also key to alleviating writing anxiety. Through asynchronous or synchronous collaboration, individuals perceive themselves as part of a collective \cite{Wang2010}. In this context, others' contributions serve not only as stepping stones for inspiration but also reduce the individual's sense of sole responsibility for the output quality through shared responsibility \cite{vass2008discourse}.

Contemporary collaborative platforms like Google Docs and Figma have advanced real-time co-editing and version control capabilities. However, these systems provide limited scaffolding for authentic co-creation dynamics such as mutual inspiration and iterative idea refinement \cite{guo2025pen,vass2008discourse}. A systematic mapping study of creativity support tools in Human-Computer Interaction (HCI) reveals that most systems prioritize individual creative processes while offering constrained support for social dynamics and team collaboration \cite{frich2019mapping}. Recent studies indicate that teams using online collaborative tools often struggle with open-ended tasks, experiencing off-topic discussions and overly divergent ideation \cite{chenyue2023online,das2024collaborative}. Research emphasizes that social interaction, communication, and group diversity remain critical yet underexplored factors in collective creativity \cite{barrett2021creative}.

Tabletop games, particularly narrative-driven ones, serve as an ancient yet vibrant social medium that integrates ``rule-based constraints,'' ``playful exploration,'' and ``collective storytelling,'' making them natural carriers of these elements \cite{cover2014creation}. \added{They help provide creators with multi-dimensional cognitive scaffolding and a low-risk space for experimentation.} \replace{They provide creators with a dynamic cognitive scaffold through game mechanics and offer a psychological buffer zone via role-playing. This enables creators to release their creativity in a low-risk social atmosphere, demonstrating significant potential for stimulating creative writing \cite{catala2012exploring}.}{Therefore, exploring how tabletop games can function as creative engines to empower creative writing has become a crucial entry point for designing novel writing support tools.}

\subsection{\replace{Tabletop}{Narrative-driven Board} Games as Creative Engines}

Tabletop games (TTGs), which refer to analogue games typically played on a flat surface and often involving physical components such as boards, cards, or tokens, have been adopted by diverse groups for purposes including entertainment, education \cite{nicholson2011making,sousa2023playing}, and social interaction \cite{gupta2025characterizing}. \replace{Given their inherent emphasis on imagination and collaborative construction, TTGs have been proven to significantly enhance participants' creative thinking, including fluency, flexibility, and originality \cite{chung2013table}, while fostering dynamic creativity experiences \cite{kuang2023memeopoly}.}{Evolving from early traditional strategy board games focused on competitive interaction to modern narrative mediums capable of carrying branching plots and role-playing, TTGs have transformed into a unique form of narrative art that emphasizes participatory experiences \cite{arnaudo2018storytelling}. In this context, narrative-driven board games, a genre where narrative elements are deeply coupled with game mechanics, demonstrate potential as creative engines \cite{sullivan2017taxonomy}. This potential is reflected not only in significantly enhancing individuals' creative thinking in terms of fluency, flexibility, and originality \cite{chung2013table}, but also in fostering a dynamic and continuously generative creativity experience \cite{kuang2023memeopoly}.}

\replace{Among these, embodied interaction is a core advantage of TTGs in supporting creative writing.}{Among these, embodied interaction, defined as how humans interact with their social and physical environments through physical movements and tangible objects, stands as a key advantage of narrative-driven board games functioning as creative engines.} Physical objects such as cards serve not only as information carriers but also as shared focal points, enabling intuitive communication through embodied actions like pointing, moving, and combining \cite{ishii2008tangible, stanton2001classroom}. This physical dimension, involving face-to-face interaction and tactile feedback, encourages broader participation, supports content manipulation, and improves creative efficiency \cite{maquil2017copse, merrill2012sifteo}. Furthermore, it allows users to offload part of their cognitive load onto the physical environment, making abstract ideas more intuitive and reducing the consumption of psychological resources \cite{Ackermann2001PiagetS}. Beyond cognitive support, \replace{TTGs}{narrative-driven board games} offer a unique sense of psychological safety \added{\cite{catala2012exploring}}. Especially in \replace{role-playing games}{narrative-driven games that emphasize role-playing}, the cost of real-world failure is suspended within this space, prompting a shift in the user's mindset from creation to ``Playfulness.'' This playfulness is crucial for fostering an encouraging environment. Moreover, through ``More-than-Human'' settings such as alien roles, users can detach from their real-world identities, thereby reducing evaluation apprehension \cite{gencc2024shroom}. Based on this dual potential in physical and psychological dimensions, \replace{many TTGs are designed as ``creative engines'' that guide creation through specific mechanical scaffolding, addressing the ``where to begin'' dilemma in creative writing}{many narrative-driven board games guide creation through various types of scaffolding \cite{sullivan2017taxonomy}, helping players break free from the dilemma of inspiration deficits}.

\replace{Specifically, these ``creative engines'' can be categorized into two types based on their functional orientation. The first type focuses on visual inspiration and intuitive association. For example, \textit{Dixit}\footnote{\url{https://mpec.it/en/tips/dixit-the-neuroscience-powered-game-that-transforms-teams/}} relies on abstract, evocative artwork to stimulate player associations; \textit{Rory's Story Cubes}\footnote{\url{https://www.rainbowresource.com/037359.html}} utilize dice with random icons as narrative anchors, guiding players to transform visual imagery into linguistic narratives. The second type emphasizes systematic support for narrative structure. For instance, \textit{Fabula Deck}\footnote{\url{https://sefirot.it/fabula-deck}} integrates a set of illustrated cards centered on core narrative elements like characters, settings, and obstacles to provide dynamic scaffolding at various stages of story development; \textit{One Hour Worldbuilders}\footnote{\url{https://kaelandm.itch.io/one-hour-worldbuilders/}} uses specific world-building prompts and strict time constraints to drive participants to rapidly synthesize heterogeneous elements and collaboratively construct coherent narrative backgrounds; \textit{Fiasco}\footnote{\url{https://medium.com/ @wendell.britt/fiasco-a-review-fb80eba4c89b}} establishes a structured, GM-less role-playing system focusing on character relationships, motivations, and dramatic conflict.}{Specifically, drawing upon Sullivan et al.'s classification of narrative board games \cite{sullivan2017taxonomy}, narrative-driven board games can be categorized into two primary types based on their degree of intervention in narrative agency. The first type focuses on image-driven association. These games provide evocative scaffolding through visual cards or discrete vocabularies without imposing systematic logical constraints, meaning that the narrative relies heavily on players' improvisation. Simple vocabulary-driven induction, such as \textit{Channel A}\footnote{\url{https://boardgamegeek.com/boardgame/134637/channel-a}} and \textit{Snake Oil}\footnote{\url{https://boardgamegeek.com/boardgame/113289/snake-oil}}, drives propositional narratives through mandatory combinations of elements. Complex visual guidance, such as \textit{Dixit}\footnote{\url{https://mpec.it/en/tips/dixit-the-neuroscience-powered-game-that-transforms-teams/}} and \textit{Rory's Story Cubes}\footnote{\url{https://www.storycubes.com/en/}}, utilizes abstract imagery to trigger cross-domain cognitive associations, whereas \textit{Once Upon a Time}\footnote{\url{https://www.rainbowresource.com/037359.html}} leverages dynamic, collaborative storytelling where players must adopt preceding narrative elements and weave plots on the fly. Meanwhile, \textit{The Quiet Year}\footnote{\url{https://buriedwithoutceremony.com/the-quiet-year}} demonstrates how minimalist prompts combined with map-drawing can trigger collective collaboration, thereby constructing an emergent community narrative.}

\added{The second type emphasizes mechanics-driven scaffolding. This category of creative engines ensures narrative coherence through systematic logical design. One form provides a logical narrative skeleton; for instance, \textit{Fabula Deck}\footnote{\url{https://sefirot.it/fabula-deck}} integrates illustrated hero's journey cards to offer step-by-step guidance, and \textit{Fiasco}\footnote{\url{https://medium.com/@wendell.britt/fiasco-a-review-fb80eba4c89b}} uses a mandatory web of character relationships to compel players to collaborate amidst conflict. These games do not predetermine plots, but instead prescribe the logical paths for creation. Another form provides a preset narrative script, such as \textit{Sherlock Holmes Consulting Detective}\footnote{\url{https://boardgamegeek.com/boardgame/2511/sherlock-holmes-consulting-detective-the-thames-mu}}. Through information exchange and collective reasoning, players piece together the narrative while exploring established clues.}

\replace{Through forced association, these games guide users to bypass logical blocks and provide initial inspiration for creative writing.}{Whether their guidance stems from card imagery or the game structure itself, these games often encourage players to establish associations between random or seemingly unrelated narrative elements.} Inspired by these structured prompting mechanisms and collaborative storytelling concepts, we developed the core framework of FIERO, a card-based creative writing system. This system deconstructs the creative process into three core phases: individual writing, group discussion, and collaborative integration. It utilizes a specialized card system as a cognitive scaffold to guide players in co-developing character settings and plot trajectories through embodied interaction\added{, ultimately creating their own heroic stories.}

\subsection{Empowerment and Challenges of AI in Collaborative Creative Writing}

Creativity can be enhanced through computing \cite{huang2023future}, with human-AI co-creation focusing on AI's role in supporting human creativity. Consequently, integrating AI technologies represents a promising direction for expanding the expressive quality of collaborative creative writing. In the field of creative writing, AI applications already span various stages, ranging from generating coherent scripts \cite{Mirowski2023} to cross-modal visual presentations \cite{jin2023generatingcoherentcomicrich}.

While collaboration can distribute creative pressure, the synergistic process within teams is not always seamless \cite{Park2023why}. AI has demonstrated \replace{unique}{distinctive} regulatory value in collaborative writing. For instance, AI can foster trust among members and improve communication efficiency \cite{HOHENSTEIN2020106190}, guiding teams toward deep strategic discussions through collaborative prompting (Co-prompting) \cite{han2024AI}. When a team becomes trapped in narrow perspectives or narrative deadlocks, AI can introduce heterogeneous viewpoints to help collective thinking transcend local maxima \cite{Johnson2025}. Although AI's real-time participation in discussions can enhance group decision accuracy, it may also pose challenges regarding the decline in team experience quality \cite{Chiang2024}. Therefore, \replace{the core of design}{a core design consideration} lies in how to leverage AI's supportive role while maintaining human creative autonomy \cite{chungTaleBrushSketchingStories2022}.

Recent studies indicate that effective AI support should enhance, rather than replace, human sense of control by providing structured prompts, reference examples, and alternative paths \cite{Dhillon2024shaping, li2024value}. In this collaborative dynamic, authors tend to view AI as a collaborative partner, expecting it to provide contextual guidance and constructive inspiration aligned with their creative intentions \cite{gero2023social}. \added{However, the potential risks of AI in creative writing cannot be ignored. Due to their probability-based prediction mechanisms, Large Language Models (LLMs) often overemphasize textual plausibility \cite{zhou2026tell}, tending to avoid sensitive topics or retain stereotypical traits to ensure discourse fluency and grammatical compliance \cite{huang2026not}. For creative writing, this safety-biased output may contribute to creative homogenization and flattened storylines \cite{anderson2024homogenization}, thereby restricting the expression of non-standard, personalized ideas by users. To counteract this algorithm-induced tendency toward mediocrity, it is worth exploring how to preserve human creative agency in AI-assisted composition.

Correspondingly, existing research has begun to integrate AI into gamified narrative practices—such as tabletop games that rely heavily on rules and scaffolding—to enhance storytelling. For instance, in a tabletop role-playing game tailored for Sámi culture, generative AI was deployed to introduce cultural and design constraints, with the AI guiding the overall story generation process to support creative expression while safeguarding narrative agency and cultural specificity \cite{roby2025storycrafting}. In another system named Memeopoly, generative AI served as a key component for for digital content, empowering creators through its ability to dynamically generate storylines in real time \cite{kuang2023memeopoly}. Furthermore, ``Trouble Maker'' demonstrated the possibility of utilizing algorithms like Markov chains to autonomously generate whimsical, physical ``Truth or Dare'' prompts, leveraging technological means to foster narrative richness \cite{fu2023trouble}. These systems robustly showcase the potential of combining generative AI with tabletop narrative workflows.}

\replace{Although many studies have explored the potential of AI in collaborative creative writing, few have situated it within the context of embodied interaction driven by card games. Currently, the HCI community still lacks an empirical understanding of how Human-AI collaboration constructs meaning within a physical tabletop game framework. Building upon the theoretical foundations of tabletop mechanics and AI's regulatory potential, we further enhanced FIERO by integrating generative AI to facilitate group negotiation and the refinement of shared narratives.}{However, most existing works focus on individual empowerment regarding creative guidance and extension. There has been limited exploration of how AI can support face-to-face negotiation, co-creation, and shared meaning-making within collocated human teams. Building upon the theoretical foundations of tabletop mechanics and AI's regulatory potential, we further enhanced FIERO by integrating generative AI to facilitate group negotiation and the refinement of shared narratives. Through the implementation and evaluation of FIERO, this paper aims to explore the benefits, challenges, and design implications of human team-AI creative writing within embodied social contexts.}


\section{Game Design}\label{sec:Designing}
\subsection{Overview of Game Design}

We present FIERO, a multiplayer web-based card game. By utilizing card game mechanics, it provides a structured framework and a social arena for storytelling. Players draw physical cards with various attributes (such as characters, locations, and opportunities) and transform them into specific narrative components like ``Heroes,'' ``Villains,'' or ``Weapons.'' Based on these elements, players engage in three rounds of structured negotiation and decision-making to co-construct a narrative. Throughout the process, the system supports collaborative writing and provides visual generation, decision assistance, and logical consistency maintenance. \replace{The ultimate goal of FIERO is to go beyond simple co-creation by lowering creative barriers and limitation}{A key goal of FIERO is to lower creative barriers and reduce creative limitations} through its integrated mechanisms, helping players craft imaginative and coherent narratives.

\subsection{Design Principle}

Mainstream collaborative tools (such as Google Docs and Figma) provide foundational support for multi-user co-editing, but their general-purpose design \replace{falls short of addressing}{may not fully address} the specific collaborative needs of creative storytelling. Therefore, this design aims to tackle key challenges in collaborative creative writing: the fragmentation of narrative threads, inefficiency in group decision-making, and a lack of confidence among creators. Accordingly, the system establishes a dedicated creative environment that integrates process guidance, decision support, and cognitive offloading. Its development is guided by the following three foundational design principles:

\subsubsection{Integrating Physical Experience and Digital Creation}

Drawing on embodied interaction theory \cite{ishii2008tangible}, this system combines the physical card game format with a digital application to leverage the strengths of both modalities. Physical cards serve as tangible social scaffolds, leveraging tactile engagement and shared visual reference points \replace{to inspire creativity and alleviate anxiety}{to help inspire creativity and reduce anxiety} in the early stages of creation \cite{chung2025toyteller}. The digital platform further supports collaborative writing and reduces cognitive friction by visualizing abstract concepts and integrating fragmented ideas, thereby \replace{overcoming the}{addressing some} limitations of purely physical or purely digital tools \cite{gupta2025characterizing}. This multimodal design aligns with cognitive load theory \cite{sweller2011cognitive}, offloading some mental tasks to external, perceptible scaffolds, allowing creators to free up cognitive resources for core narrative ideation.

\subsubsection{Intelligent Assistance Preserving Human Creative Agency}

Consistent with human-AI collaboration research \cite{gero2023social}, the intelligent assistance in the system is positioned as a supportive collaborator rather than a directing ``director.'' Its core function is to assist in resolving common coordination challenges during collaboration, such as conflicting ideas or breaks in narrative logic, while respecting and maintaining the final creative control of human authors. This design \replace{avoids}{seeks to mitigat} the risk of conceptual convergence that may arise from excessive AI intervention \cite{biermann2022tool}, preserving the social dynamics essential for collaborative creativity \cite{barrett2021creative}. Specifically, when the team reaches a decision impasse, the system can select the most plausible plot development option based on the context, providing reasoning to serve as an ``icebreaker'' without imposing a single direction \cite{han2024AI}. It can also identify thematic links and causal relationships between different user contributions and suggest organic integration strategies \cite{chungTaleBrushSketchingStories2022}. Furthermore, the system offers text-polishing support while preserving the author's personal style \cite{brade2023promptify}.

\subsubsection{Balancing Guidance and Freedom with Flexible Narrative Structure}

To alleviate the confusion and pressure creators face at the outset, often expressed as ``not knowing where to start,'' the system introduces a flexible narrative framework \cite{karwowski2018measuring}. This framework provides foundational structure through distinct card categories and three narrative rounds, while maintaining support for open-ended storytelling. Inspired by research on constrained creativity \cite{han2024AI}, this design uses a moderate degree of structure to reduce cognitive load without stifling originality. Specifically, the cards feature a visual style that balances suggestive imagery with ample abstraction, enabling players to generate diverse narrative possibilities through multi-card combinations. This approach accommodates both users who prefer guided structure and those who favor free expression, \replace{effectively}{potentially} lowering the participation barrier for collaborators of varying experience levels \cite{frich2019mapping}.

\subsection{Gameplay}
\subsubsection{Game Flow}

The game is designed for a three-player collaborative writing scenario, consisting of three rounds corresponding to the beginning, confrontation, and resolution of the story. The overall workflow is illustrated in Figure \ref{fig: Interaction}, with the detailed steps as follows:

\begin{figure}[t]
\centering
\includegraphics[width=\linewidth, height=0.8\textheight, keepaspectratio]{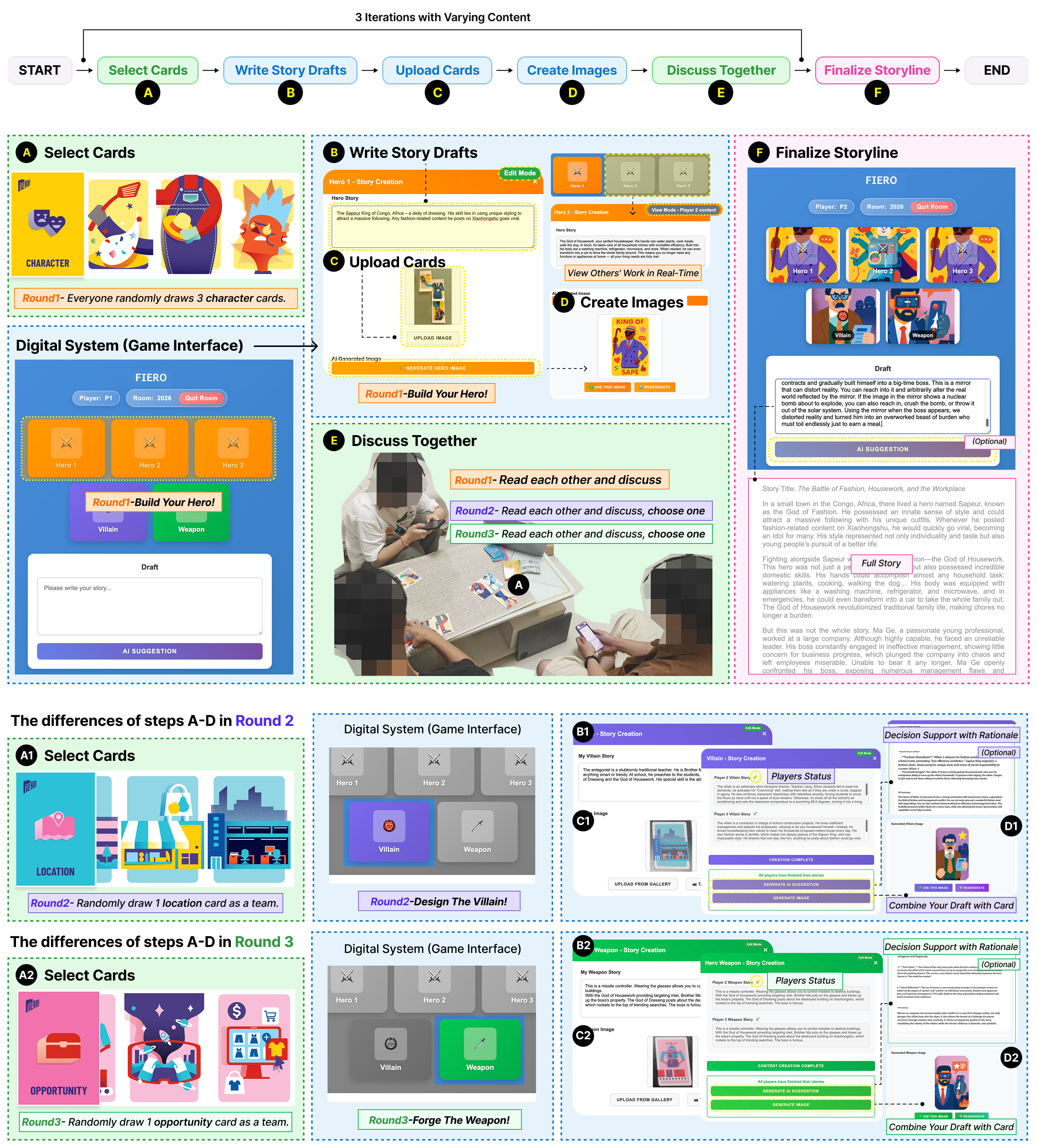}
\caption{FIERO Game Flow: (A) Players randomly draw physical cards; (B) Based on the cards, they collaboratively write a story within the system; (C) Upload the drawn cards into the system; (D) New cards are generated based on the cards and story content; (E) Discuss their own stories and subsequent developments with teammates; (F) After three rounds of writing, collaboratively integrate and polish the final story.}
\label{fig: Interaction}
\end{figure}

\textbf{Round 1: Hero Creation (Beginning)}
\begin{itemize}
\item \textbf{Draw Cards}: Each player draws three Character cards from the table (Figure \ref{fig: Interaction} - A).
\item \textbf{Creation}: In the system, select the ``Hero Frame'' and independently create a superhero character along with their backstory within the document editing area (Figure \ref{fig: Interaction} - B).
\item \textbf{Integration}: Players can take photos of the drawn cards and upload them to the system (Figure \ref{fig: Interaction} - C), combining them with the written text to generate new character image cards (Figure \ref{fig: Interaction} - D).
\item \textbf{Review and Discussion}: During creation, players can switch to view others' hero frames to track progress. After completion, each player introduces their character and story, with details supplemented through group questioning (Figure \ref{fig: Interaction} - E).
\end{itemize}

\textbf{Round 2: Villain and Scene Setting (Confrontation)}
\begin{itemize}
\item \textbf{Draw Cards}: All players collectively draw one Location card (Figure \ref{fig: Interaction} - A1).
\item \textbf{Collaborative Creation}: Enter the system's ``Villain Frame'' and, based on the three created heroes, independently develop a villain character along with an associated scene story (Figure \ref{fig: Interaction} - B1). Players can view each other's edits in real-time during creation.
\item \textbf{Integration}: After creation, upload a photo of the Location card (Figure \ref{fig: Interaction} - C1) and combine it with the text to generate a villain image card (Figure \ref{fig: Interaction} - D1).
\item \textbf{Discussion and Decision-making}: Each player presents their villain story. The group then discusses and agrees on the most cohesive storyline (Figure \ref{fig: Interaction} - E). If consensus cannot be reached, the system's built-in decision support function (Figure \ref{fig: Interaction} - D1) can assist in arbitration.
\end{itemize}

\textbf{Round 3: Weapon and Climax Setting (Resolution)}
\begin{itemize}
\item \textbf{Draw Cards}: All players collectively draw one Opportunity card (Figure \ref{fig: Interaction} - A2).
\item \textbf{Collaborative Creation}: Enter the system's ``Weapon Frame'' and, incorporating the existing hero and villain stories, independently create a super-weapon along with a key event in the plot (Figure \ref{fig: Interaction} - B2). Players can view each other's edits in real-time during creation.
\item \textbf{Integration}: After creation, upload a photo of the Opportunity card (Figure \ref{fig: Interaction} - C2) and combine it with the text to generate a weapon image card (Figure \ref{fig: Interaction} - D2).
\item \textbf{Discussion and Decision-making}: Each player presents their weapon story. The group then discusses and selects the most suitable one to integrate into the main plot (Figure \ref{fig: Interaction} - E). Decision support (Figure \ref{fig: Interaction} - D2) is available again if needed.
\end{itemize}

\textbf{Final Stage: }Story Synthesis and Polishing
All players enter the system's public editing area to collaboratively write the complete story. Content generated in the previous three rounds can be reused, or players can freely create new content. Narrative suggestions (Figure \ref{fig: Interaction} - F) are available throughout the editing process.

\subsubsection{Physical Card Design}

\begin{figure}[t]
\centering
\includegraphics[width=\linewidth, height=0.8\textheight, keepaspectratio]{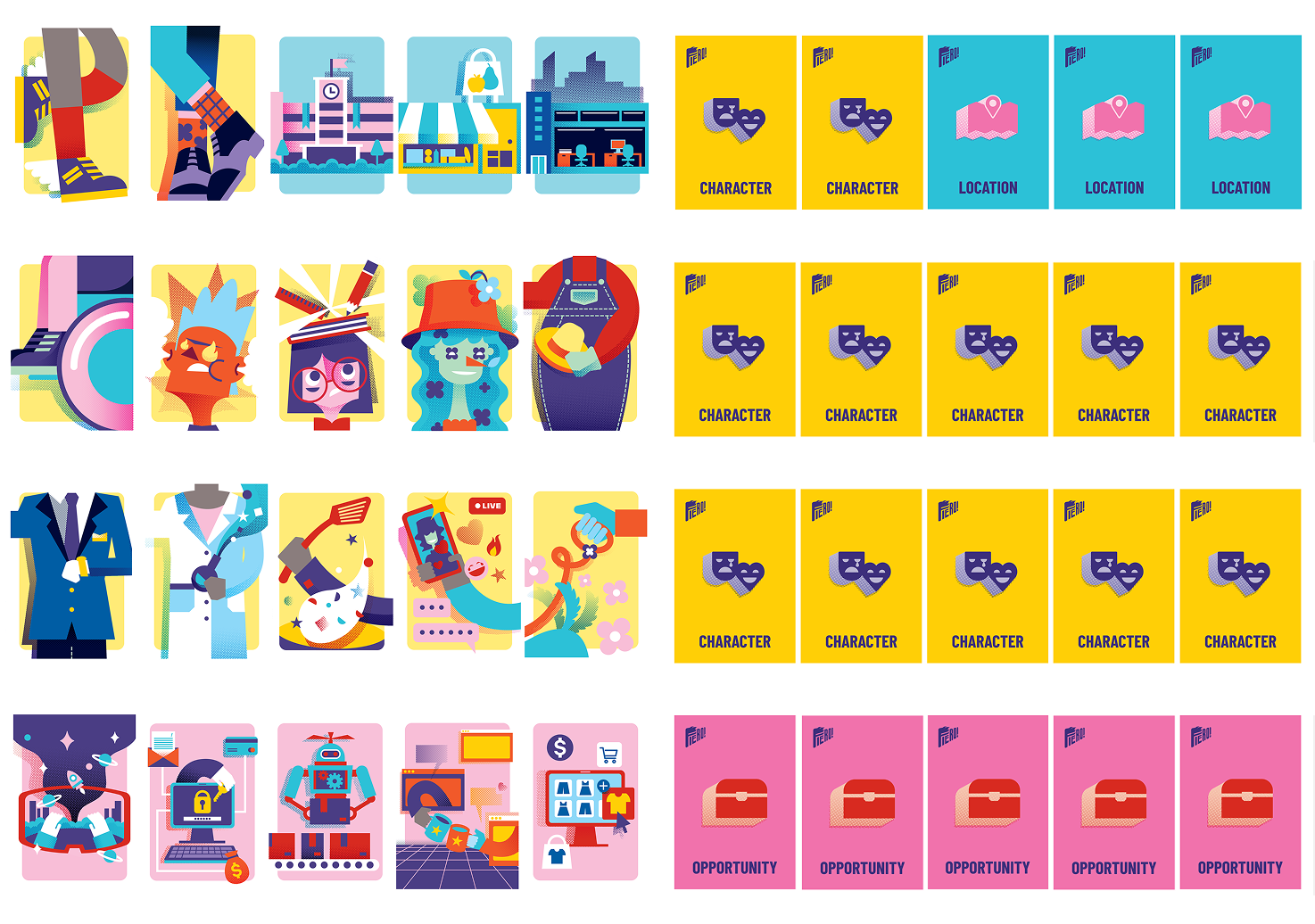}
\caption{The game includes 20 physical cards in three categories: Character cards (yellow, 12 cards) depicting modular body parts (head, hands, body, limbs); Location cards (blue, 3 cards); and Opportunity cards (pink, 5 cards).}
\label{fig:Card}
\end{figure}

The physical cards of this game are designed as key cognitive scaffolds to support narrative construction. Their three main categories inherently correspond to the classic three-act narrative structure (beginning, confrontation, and resolution), providing a clear procedural framework for collaborative story creation \cite{cover2014creation}. The card design is illustrated in Figure \ref{fig:Card}.

\begin{itemize}
    \item \textbf{Character Cards: }As the beginning of the narrative, Character cards guide players in creating superhero characters. Each card depicts a character part or core feature (e.g., head, torso, hand, foot) accompanied by abstract visual cues. This design provides a structured starting point while preserving considerable customizability for character backgrounds and abilities. By segmenting character parts, the game increases combinatorial freedom (e.g., allowing fantastical combinations like a ``two-headed'' character) and offers specific inspirational anchors (e.g., a ``fire-breathing eye'' card can inspire corresponding superpower backgrounds). Additionally, the design does not force players to assemble a complete character, thereby maintaining narrative openness for character depth and relationship development.
\end{itemize}

\begin{itemize}
    \item \textbf{Location Cards: }As the focus of the confrontation stage, Location cards guide the team in co-creating the villain and their story. The cards depict settings with distinct environmental attributes. Their function is to provide a stage for the villain's motivations and serve as a catalyst for plot conflict. They establish the necessary spatial, temporal, and logical framework for the story, enabling players to conceive the villain's plans, schemes, or origin story around the setting, thereby facilitating team consensus on the core conflict.
\end{itemize}

\begin{itemize}
    \item \textbf{Opportunity Cards: }As the turning point toward the resolution, Opportunity cards guide the team in creating key super-weapons and driving the story's climax and conclusion. The cards contain more dynamic and interactive visual elements (e.g., conveyor belts with robots, screens with mice), aiming to provide the hero team with crucial tools or opportunities to break deadlocks and confront the villain. This type of card encourages creative problem-solving among participants, steering the narrative toward a transformative and resolving climax that naturally leads to the story's conclusion.
\end{itemize}

Visually, the cards employ an abstract style to avoid limiting players' imagination, while their physical form supports intuitive social interaction (such as pointing, arranging, passing), enhancing the team's shared reference and situational resonance \cite{stanton2001classroom}.

\subsubsection{Digital Interaction Interface}

This interface serves as the central hub integrating physical and digital interaction. Its core design consolidates all story elements (heroes, villains, weapons) into independent visual containers, paired with a shared document editing area (Figure \ref{fig: Interface}), allowing players to maintain a comprehensive overview of the narrative while co-writing. The interface primarily consists of the following core components:

\begin{figure}[t]
\centering
\includegraphics[width=1\textwidth]{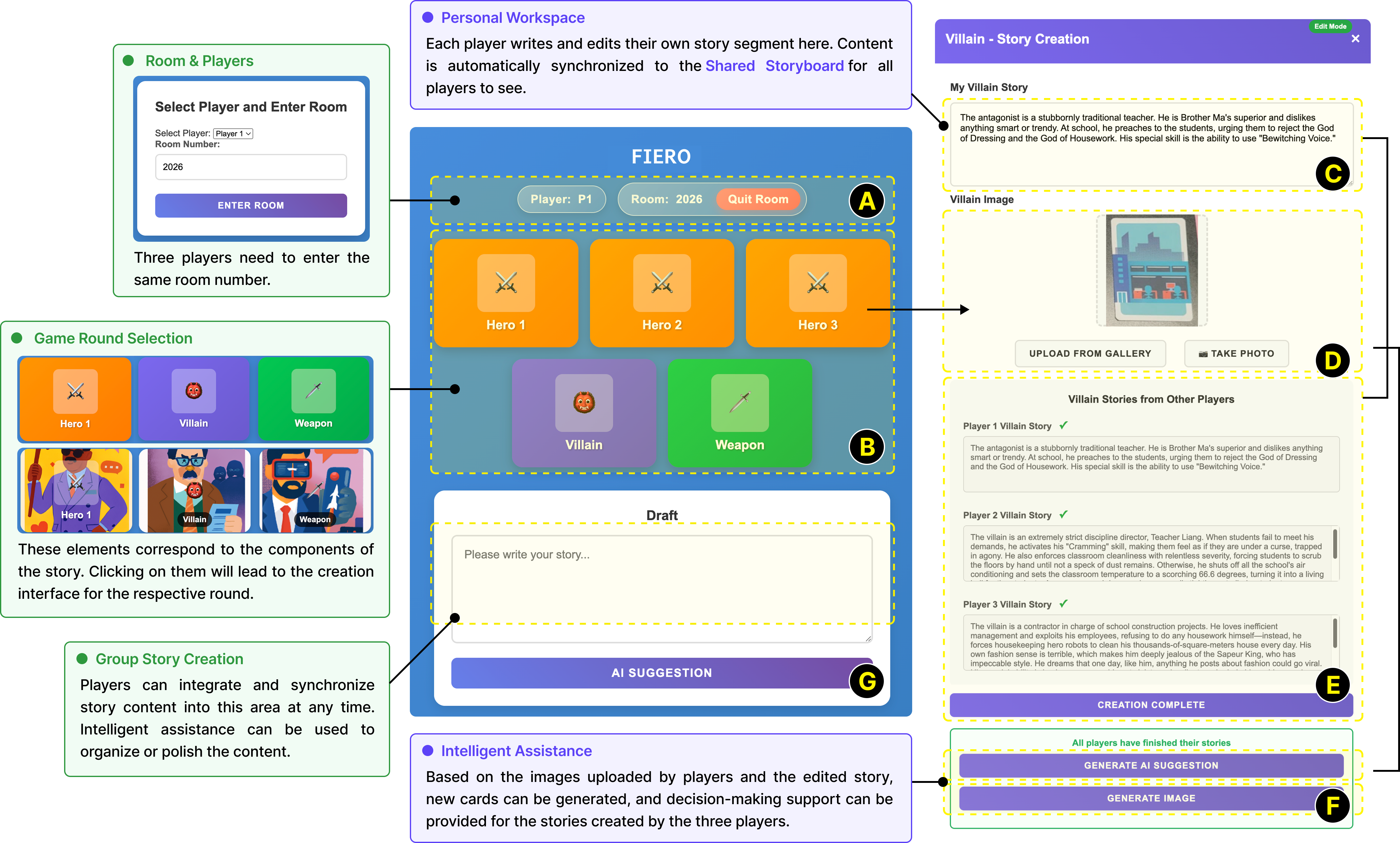}
\caption{Digital Interface: (A) Players can join a game by selecting the corresponding room number to form a team; (B) Players can enter the story creation phase following a structured workflow; (C) After clicking (B), players can write in this area; (D) Card upload area; (E) Real-time view of other players' writing progress; (F) Access system assistance, including decision-making suggestions and image generation; (G) Area where all players collaboratively write the final story.}
\label{fig: Interface}
\end{figure}

\begin{itemize}
    \item \textbf{Room \& Players:} This area displays the current room ID and player information, serving as the foundational entry point for multi-user collaborative creation. Three players must enter the system by inputting the same room code (Figure \ref{fig: Interface}A).
\end{itemize}

\begin{itemize}
    \item \textbf{Game Round Selection:} The three rounds of creation (hero, villain, weapon) are modularly presented here, also serving as a visual index for the cards required for the complete story. Players click these modules to enter the creation interface and complete the corresponding section (Figure \ref{fig: Interface}B). Created character, villain, and weapon images are displayed here, forming shared visual anchors to enhance narrative coherence \cite{masson2024visual, qin2024charactermeet} and providing a global perspective on the story's development.
\end{itemize}

\begin{itemize}
    \item \textbf{Personal Workspace:} After entering the creation interface, players can draft their stories within a personal editing box (Figure \ref{fig: Interface}C). Upon completion, the content is automatically synchronized with other players (Figure \ref{fig: Interface}E), enabling real-time viewing and reference, fostering transparency in the creative process, mutual inspiration, and facilitating discussion.
\end{itemize}

\begin{itemize}
    \item \textbf{Intelligent Assistance:} When all three players complete their individual creations and enter the group discussion phase, they can access intelligent assistance if encountering narrative disagreements or difficulties in progression (Figure \ref{fig: Interface}F). Based on the existing story context, the system generates the most plausible subsequent plot options with explanations, thereby aiding the team in making efficient decisions and breaking discussion deadlocks. After players upload their drawn cards (Figure \ref{fig: Interface}D), the system \replace{can combine the player's story from the personal workspace to generate a new card image}{performs an image-to-image generation by combining the inherent visual features of the card with the story text authored by the player in the personal workspace, thereby generating a synthesized new card image} (Figure \ref{fig: Interface}F).
\end{itemize}

\begin{itemize}
    \item \textbf{Group Story Creation:} This is a collaborative, real-time text editor. In this area, the team can integrate various narrative elements from the three independent rounds, jointly write, revise, and ultimately complete the full story narrative (Figure \ref{fig: Interface}G). When necessary, intelligent assistance can provide suggestions for paragraph optimization, style unification, or logical coherence, aiding the team in systematically polishing the complete story.
\end{itemize}

\subsection{Implementation}

We adopt a Client-Server architecture and leverage modern web technology stacks to develop an online game application supporting real-time multi-user collaboration. The frontend is constructed using native HTML, CSS, and JavaScript, implementing a dynamic Single-Page Application (SPA) user interface. The frontend interface is divided into two main components: an entry page for player selection and room access, and a main creation page containing multiple cards (heroes, villains, weapons) along with a shared editing area. By clicking on different cards, users can open corresponding modals to conduct independent story and image creation.

The client establishes a WebSocket connection with the backend server using the Socket.IO\footnote{\url{https://socket.io/}} client library to handle functionalities such as room status management and text box input. \replace{For the text generation functionality via LLM and image generation functionality, we employed the GPT-4 and GPT-4o image models respectively, with specific prompt engineering detailed in Appendix \ref{appendix:prompts}. When users click the corresponding buttons (e.g., generate image, generate suggestion buttons), the frontend invokes the API and stores the generated content on the server.}{For in-game text generation we employ GPT-4 (decision support: $\textit{max\_tokens}{=}1000$, $\textit{temperature}{=}0.7$; final-stage story suggestion: $\textit{max\_tokens}{=}800$, $\textit{temperature}{=}0.7$), and for image generation we employ the GPT-4o image model invoked through an OpenAI-compatible chat-completions endpoint (aspect ratio $1{:}1$). Each AI call is stateless from the model's perspective: the server assembles the current room state—all player drafts of the present and prior rounds—into a single user prompt and combines it with a role-conditioned system prompt before invocation. Verbatim system prompts, user-prompt assembly templates, and decoding parameters are provided in the Appendix \ref{appendix:prompts}. The complete implementation of the proposed framework is open-source and accessible on GitHub\footnote{\url{https://github.com/zxioke/FAIero-game}}.}

\subsubsection{A Three-Layer Pipeline for Resolving Concurrent Input Conflicts}

\added{Because three players independently author hero, villain, and weapon drafts before each round's discussion phase, the model typically receives three different drafts as input. FIERO resolves this through a three-layer pipeline rather than AI arbitration: (i)~\emph{shared cards} constrain the input space prior to LLM invocation---e.g., a single Location card collectively drawn in Round~2 anchors all three villain drafts to a common setting; (ii)~the LLM is constrained by its system prompt to \emph{select among player-authored drafts} rather than invent new content, and to surface its rationale as plain text in the suggestion panel; (iii)~the subsequent face-to-face discussion is where the group accepts, modifies, or overrides the AI's choice. The shared storyboard does not lock to the AI's selection. To verify that this pipeline holds under stress, we conducted a pre-deployment adversarial evaluation covering hard contradictions, genre and tonal conflicts, underspecified or empty drafts, transgressive content, and prompt-injection attempts embedded in player drafts; details are reported in the Appendix \ref{appendix:adversarial}.}

\section{User Study}\label{sec:UserStudy}
To investigate how assisted collaborative writing gameplay influences players' engagement in co-constructing stories (RQ1) and how the collaborative stories produced in this gameplay experience differ from those produced through collaborative writing alone (RQ2), we conducted a between-subjects experiment comparing a game-based collaborative writing condition with a game-free collaborative writing baseline condition.

\subsection{Study Design}
In this between-subjects experiment, 60 participants were randomly assigned to either a story creation group or a card game-based story creation group. Participants were organized into groups of three to complete the experiment. Both groups followed the same experimental procedure: creating a superhero story within 30 minutes. The experiment was divided into three rounds, each consisting of 5 minutes of individual thinking followed by 5 minutes of group discussion. The detailed procedure was as follows:

\begin{itemize}
    \item \textbf{Round 1}: Each participant independently created a superhero character and develops a background story. Afterward, they introduced their characters to the group.
    \item \textbf{Round 2}: Each participant continued to independently create a story scene and a villain character based on the three superheroes from the team. The group then discussed and agreed on the most appropriate scene direction.
    \item \textbf{Round 3}: Each group member created a superhero weapon based on the previous two rounds' content and selected the most reasonable weapon. Finally, the group discussed and integrated the content from the previous rounds to finalize the story.
\end{itemize}

\replace{The experimental process is the same for both the Story Creation Group and the Card Game Based Story Creation Group, but the tools used differ.}{Both the Story Creation Group and the Card Game Based Story Creation Group adopted a three-person, face-to-face collaborative communication format. The two groups remained entirely identical in terms of their overall timeframe, round progression, and narrative tasks, but participants utilized different interaction mediums within the same narrative phases.} The specific procedures for each group were as follows and shown as Figure \ref{fig:placeholder}:

\begin{figure}
    \centering
    \includegraphics[width=1\linewidth]{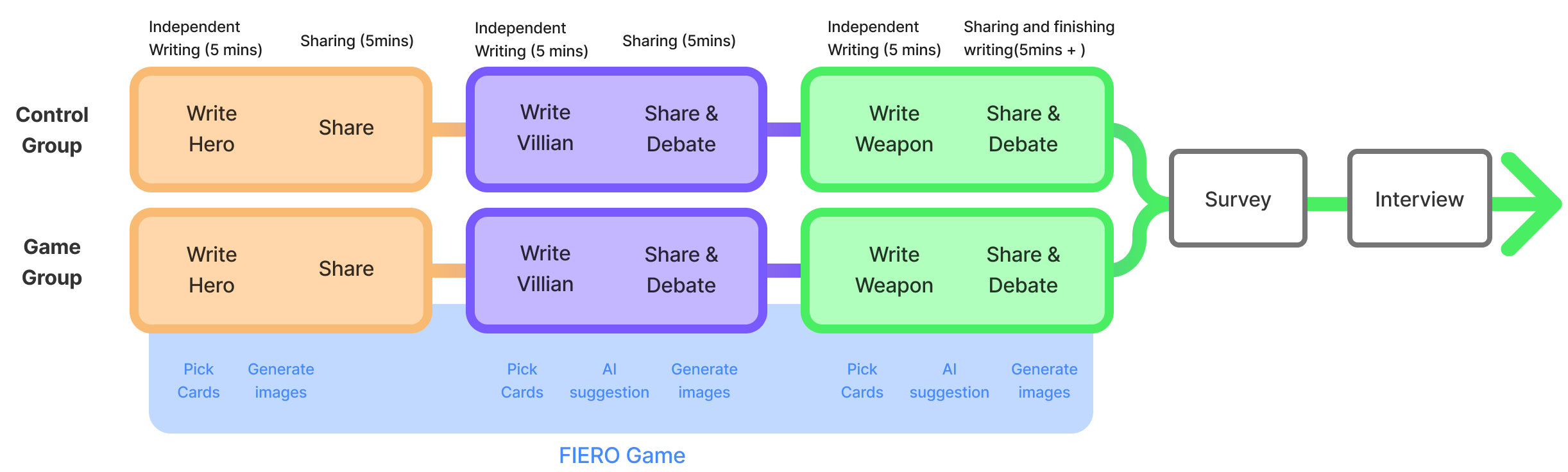}
    \caption{Experimental Procedure for the Control and FIERO Groups.}
    \label{fig:placeholder}
\end{figure}

\begin{itemize}
    \item \textbf{Control Group (Story Creation Group): referred to as ``C''}: \replace{The Story Creation Group does not use cards but instead follows written prompts provided by the experimenter. For each round, participants create stories based on the given prompts. The members of the Story Creation Group collaboratively document their creative process and final story in a shared Google Doc.}{The three participants in the Story Creation Group co-utilized a single shared Google Docs page. During the independent ideation phase of each round, participants reflected based on the tasks provided on-site (see Appendix \ref{appendix:Written prompts}) and individually recorded their initial ideas within the same page of the document. During the negotiation phase of each round, group members engaged in face-to-face consultation centered around the synchronized text page, made decisions on the narrative direction through mutual text review within the document and oral debates.}
    \item \textbf{FIERO Group (Card Game Based Story Creation Group): referred to as ``G''}: \replace{The Card Game Story Creation Group follows the process outlined by the card game. Participants draw various types of cards (such as characters, scenes, weapons, etc.) to build and develop the story. The Card Game Story Creation Group also documents their creative process and final story in the game.}{The three participants in the Card Game Story Creation Group entered the same game room page. During the independent ideation phase of each round, participants reflected based on the tasks provided on-site (see Appendix \ref{appendix:Written prompts}), introduced narrative prompts by drawing physical cards (such as characters, scenes, weapons, etc.), and individually recorded their initial ideas in the corresponding sections of the game system. During the negotiation phase of each round, group members engaged in face-to-face consultation centered around the synchronized game page, which encompassed story text, rendered visuals, and optional AI suggestions, made decisions on the narrative direction through mutual system review and oral debates.}
\end{itemize}

Before the formal experiment began, all participants received detailed written and verbal instructions outlining the game objectives, gameplay rules, and expected writing tasks. Each group completed a short practice session to familiarize themselves with the game mechanics and collaborative process. The specific procedural instructions are provided in the Appendix \ref{appendix:Written prompts}.

With participants' informed consent, all gaming sessions and interviews were audio and video recorded. The researchers also took observational notes of participants' interactions, decision-making behaviors, and writing activities, which were cross-referenced with the audio and video recordings for subsequent analysis. Immediately after each gaming session, participants were asked to complete a post-experiment questionnaire assessing their creativity support index, creative self-efficacy, and story satisfaction. Each group also participated in a semi-structured interview to reflect on their experiences with the collaborative writing process. All data were anonymized and transcribed for qualitative and comparative analysis.

\subsection{Quantitative Data}
\subsubsection{Questionnaires}

After completing the collaborative writing task, each group of participants first filled out a questionnaire to investigate their subjective experiences during the creative process. All measurement items were rated on a 7-point Likert scale (1 = Strongly Disagree, 7 = Strongly Agree). The specific measurement tools included the Creativity Support Index, the Creative Self-Efficacy Scale, and the Story Satisfaction Assessment. Detailed items are provided in the Appendix \ref{app:Questionnaire}.

\textbf{The Creativity Support Index (CSI)}, developed by Cherry and Latulipe \cite{cherry2014quantifying}, is a standardized scale specifically designed to evaluate the extent to which a tool or environment supports creative activities \cite{carroll2009creativity}. The original CSI comprises six dimensions. For this study, the items were adaptively modified to suit the context of collaborative story creation, and four dimensions were employed to specifically assess the experiential aspects most relevant to our creative storytelling environment: Expressiveness, Immersion, Appreciation of Outcomes, and Tool Transparency. The ``Exploration'' dimension was not included because content related to idea generation was already measured by the subsequent Creative Self-Efficacy Scale. The ``Collaboration'' dimension was not included because both experimental conditions involved collaborative writing; the study compared ``different modes of collaboration.''

\textbf{Creative Self-Efficacy (CSE)} refers to an individual's belief in their ability to perform creative tasks successfully. This study employed a scale developed by Tierney and Farmer \cite{haase2018meta} to measure participants' creative confidence throughout the process of idea generation, idea development, and idea execution during collaborative writing. Our research adapted the scale to fit the specific context of collaborative story creation in our experimental setting, resulting in a total of five items. These items encompass aspects of creative fluency, novelty, originality, creative extension, and creative persuasion. These items enable us to measure participants' confidence in their creative performance, both cognitively and socially, within team-based storytelling activities.

Participants' evaluation of their collaborative outcomes (\textbf{Story Satisfaction Assessment (SSA)}) was measured using a scale adapted based on the Transportation Theory \cite{green2000role}. This scale assesses participants' satisfaction with the final story from three dimensions: Narrative Quality, Collaborative Process, and Individual Contribution. Our final instrument consists of five items, covering narrative logic, narrative creativity, collaborative synergy, narrative integration, and creative autonomy. This assessment provides a theoretically grounded, multi-dimensional perspective on participants' evaluations of the final story.

\subsubsection{Content analysis}

We employed \textit{LLM as a Judge} to evaluate the final writing outcomes of both the FIERO group and the control group. LLM as a Judge refers to the use of a large language model to systematically assess and compare textual outputs based on predefined criteria. This approach leverages the model's capacity to provide consistent and scalable judgments, reducing subjective bias that may arise from human evaluators \cite{kim2025evaluating}. Following previous research \cite{kim2025codi}, we evaluated the generated stories across seven dimensions:

\begin{itemize}
    \item \textbf{Plot Coherence}: Evaluates story arc completeness, narrative flow, plot development logic, conflict setup and resolution, and pacing. Stories with clear beginning-middle-end structure, logical plot progression, well-developed conflicts with satisfying resolution, and smooth narrative flow are preferred over those with confusing or incomplete plots, illogical progression, unresolved conflicts, or abrupt transitions.
    
    \item \textbf{Development}: Assesses world-building depth, character development, setting details and richness, story progression, and background information. Stories with rich and detailed world-building, well-developed characters with clear backgrounds, vivid setting descriptions, and gradual story progression are preferred over those with shallow world-building, underdeveloped characters, vague setting details, or rushed progression.
    
    \item \textbf{Language Use}: Examines writing style sophistication, vocabulary richness, sentence structure variety, clarity of expression, expressiveness, and readability. Stories with sophisticated writing style, rich and appropriate vocabulary, varied sentence structures, and clear expressive language are preferred over those with simple or repetitive writing style, limited vocabulary, monotonous sentence structures, or unclear expression.
    
    \item \textbf{Anthropomorphism}: Measures character depth and complexity, human-like qualities (emotions, thoughts, motivations), relatability, emotional expression, and psychological realism. Stories with deep and complex characters, rich human-like qualities, highly relatable characters, and vivid emotional expression are preferred over those with shallow or one-dimensional characters, lack of human-like qualities, or unrealistic mechanical portrayal.
    
    \item \textbf{Character Fidelity}: Evaluates character consistency throughout the story, staying true to established character traits, believable behavior, character coherence, and absence of contradictions. Stories with consistent character portrayal, characters staying true to their established traits, and believable coherent behavior are preferred over those with inconsistent portrayal, characters acting out of character, or contradictions in character actions or personality.
    
    \item \textbf{Creativity}: Assesses originality of ideas, innovative concepts, creative elements, uniqueness, fresh perspectives, and imaginative content. Stories with highly original ideas, innovative concepts, creative and unique elements, fresh perspectives, and highly imaginative content are preferred over those with common or clichéd ideas, lack of innovation, uncreative conventional elements, or unimaginative content following familiar patterns.
    
    \item \textbf{Overall Preference}: Reflects overall story quality, reader engagement, enjoyment factor, completeness, and coherence across all aspects. Stories with higher overall quality, more engaging content, more enjoyable reading experience, more complete narrative, and better coherence across all aspects are preferred over those with lower overall quality, less engaging content, or poor coherence across aspects.
\end{itemize}

\subsection{Qualitative Data}

We conducted semi-structured interviews with all participants to gain a deeper understanding of how they interpreted the collaboration mode during the writing process, as well as the role played by the collaborative writing tools used. The interviews specifically inquired about the ways participants drew inspiration during the collaborative process. All interview data were audio-recorded, transcribed, and anonymized with participants' consent, and were subjected to thematic analysis. The outline of the semi-structured interviews is detailed in the Appendix \ref{app:Interview}.

Furthermore, participants' interaction behaviors during the experiment were comprehensively recorded, including the written content from each round, the actions taken during the communication process, and the discussions that took place. Finally, the stories completed by each team also served as material for content analysis.

\subsection{Participants}

\added{We recruited participants who were interested in collaborative creative writing but did not consider themselves proficient in creative writing, through word-of-mouth and online recruitment via social media.} We recruited 30 participants for each of the Control and FIERO groups (divided into 10 teams of 3 members each) to take part in an offline collaborative writing activity. The study cohort consisted of 25 males (42\%) and 35 females (58\%), with ages ranging from 18 to 28. Among these 60 participants, our team compositions included varied gender distributions: all-male teams, all-female teams, mixed teams with two males and one female, and mixed teams with two females and one male. \added{The majority of participants were students (80\%), with diverse academic backgrounds spanning design, science and engineering, humanities and social sciences, and business. The remaining non-student participants (20\%) came from fields such as technology, design, media production, marketing, and administrative office work. Detailed participant information is provided in Appendix \ref{app:Participants}. None of the participants had professional publishing or screenwriting backgrounds or prior experience in creative writing. All were non-professional creators.} This study was approved by the Institutional Review Board and conducted ethically in accordance with university IRB regulations. All participants consented to the anonymous use of their data.

\subsection{Data Analysis}

To address the research questions: how people engage in collaborative writing (RQ1) and how the stories created under the two conditions differ (RQ2), we employed a multi-dimensional measurement approach, systematically collecting both quantitative and qualitative data to comprehensively evaluate participants' engagement processes, creative performance, and collaborative outcomes under the two collaborative writing conditions. The data collected for this study included quantitative questionnaire responses and writing outputs from participants in both conditions, as well as qualitative data from their interaction processes and semi-structured interviews.

\subsubsection{Questionnaire Analysis}

For the questionnaire data, we conducted Mann-Whitney U tests to compare the differences in scores between the experimental group and the control group on the Creativity Support Index, Creative Self-Efficacy Scale, and Story Satisfaction Assessment (see Appendix \ref{app:Questionnaire} for details). Concurrently, Cronbach's alpha was calculated for each scale to evaluate the internal consistency of the measurement items, with all coefficients exceeding 0.763, indicating acceptable reliability. \added{Additionally, effect sizes ($r$) were calculated as $r = Z/\sqrt{N}$ ($N = 60$) and interpreted following Cohen's conventions ($r = 0.1$, $0.3$, and $0.5$ as small, medium, and large effects, respectively). 

To conduct a more granular analysis, we further compared 15 individual items (item-level) across the three scales. Because multiple comparisons inflate the probability of false-positive findings, we applied a Bonferroni correction separately to each family of tests, resulting in an adjusted significance threshold of $\alpha = 0.05/15 = 0.00333$. While this strict correction effectively controls for Type I errors, its highly conservative nature increases the risk of Type II errors (false negatives). Under this adjusted threshold, only item SSA2 remained statistically significant ($p = 0.003 < 0.00333$). Consequently, other nominally significant results ($p < 0.05$, uncorrected) should be interpreted with caution and viewed primarily as exploratory or hypothesis-generating.} 

\subsubsection{Content Analysis}

We employed three recent large language models, Claude Opus 4.5, Gemini 3 Pro, and GPT 5.2, which have not been used in the FIERO system, to analyze the results. Using multiple independent evaluator models allowed us to reduce potential model-specific preference bias and improve the robustness of the evaluation. For each comparison, we adopted a pairwise evaluation approach \cite{kim2025evaluating}: each of the nine texts from the FIERO group was compared with each of the nine texts from the Control group. To mitigate order effects, we conducted each comparison in both AB and BA directions, outputting only a win or loss for each round. If the two rounds yielded inconsistent outcomes for a text pair, the result was recorded as a draw. The prompts used for all three models were identical and co-designed by an industry practitioner with five years of experience in game narrative and a master's degree in game design (see Appendix \ref{app:content} for details).

\subsubsection{Thematic Analysis}

We employed an iterative, consensus-based thematic analysis approach \cite{hammer2014confusing} on the post-experiment interview data. After transcription, three researchers independently conducted open coding to identify themes related to writing interactions and experiences. Through collaborative discussions and preliminary exercises, we developed an initial codebook, regularly refining the codes to ensure reliability until theoretical saturation was achieved. The initial coding covered the following aspects: perceptions of the writing method (e.g., ``someone needs to lead and drive the process,'' ``cards help stimulate imagination''), interaction behaviors during the creative process (e.g., ``pointing at cards during discussions,'' ``looking at each other's phones,'' and ``offline eye contact and facial expressions''), reflections on creative outcomes (e.g., ``inspiration came from life,'' ``difficulty integrating different worldviews''), and emotional experiences during the creative process (e.g., ``enhanced social aspect,'' ``sense of immersion''). Finally, by examining contextual meanings and eliminating irrelevant codes, we iteratively refined these codes into core themes.

\subsubsection{Behavioral Observation}

During the experiment, researchers recorded participants' non-verbal interaction behaviors during collaborative writing, such as pointing at cards, looking at each other's phone screens, and gestural communication. These observations provided supplementary contextual information for understanding participants' collaboration patterns and were used for probing and validation during the interviews.

\section{Result}\label{sec:Result}
\subsection{Player Engagement in Assisted Collaborative Storytelling}

To evaluate how players used assisted tools to co-construct stories in collaborative creation games (RQ1), we observed and analyzed each group's works. FIERO expanded interactive space and decision-making, encouraging deeper communication and collaboration. This led to more logically coherent story plots and strengthened emotional and strategic interactions between characters (see Section \ref{sec:6.1.1}). In contrast, while the stories from the Control group were creative, the character interactions remained relatively shallow. They lacked emotional depth and strategic decisions, resulting in simplistic structure. Furthermore, FIERO broke away from the traditional hero model. It helped players create more modern and innovative characters and plots, avoided the simplistic ``superpowers vs. villains'' framework (see Section \ref{sec:6.1.2}).

\subsubsection{FIERO Makes Stories More Rigorous and Hierarchical}\label{sec:6.1.1}\

In story content, the Control group focused more on individual development than detailed integration of the three heroes. For instance, the collaboration among characters in C4 remained based on task completion and complementary abilities \textit{(evacuating super citizens in super Guangzhou, clearing the mind-control facilities of the Light Energy Race, and defeating their supervisor during a new invasion)}. However, their emotional and intellectual interactions were rather shallow \textit{(during an alien resistance operation, a super city fell, and the three heroes—who happened to be operatives in the mission—crossed paths while retreating)}. Characters functioned independently; collaboration was mostly combat coordination, lacking emotional exchange or detailed strategies. Despite interesting elements like \textit{aliens, a ``super Earth,'' and superpowers}, the plot lacked logical structure and was shallow. For example, in C4's story, \textit{``after Character X escapes from a laboratory, they encounter other heroes and collaborate to fight the aliens, but this interaction is simplistic and lacks in-depth strategic decision-making: after sending a retreat signal, the three heroes receive no response from the super Earth. They then struggle to survive in a city overrun by the Light Energy Race, eventually meeting and cooperating to stay alive. After successfully killing a Light Energy Race supervisor, they discover a connection between the Race and the super Earth's Ministry of Truth on his body.''} 

With FIERO's assistance, story progression and character interactions became more complex and multidimensional. Through visualization technology, plots could be flexibly adjusted according to context, and characters' decisions balanced personal goals with team needs. In G7's story, for example, \textit{``hero Ollie's goal was not merely to shop but to meet heroes Dr. Hua and Bubble—with whom they were unfamiliar regarding each other's superpowers.''} The trio's gathering had clear causes and consequences: \textit{``upon getting acquainted, Dr. Hua realizes Ollie was once her colleague, and Bubble reveals the enemy's ability to manipulate fruit coloring. The three decide to join forces against the villain, successfully using their respective abilities to subdue them in an alternate world and ultimately securing Dr. Hua's laboratory data.''} In G6's story, heroes merged into a robot: \textit{``Hero C controls the brain to command Heroes A and B, with Hero A acting as the attack unit using flame-throwing lasers to pursue the villain, and Hero B handling internet connectivity, crowd evacuation, and logistics.''} This subordination of individual prowess to collective strategy exemplifies FIERO's layered decision-making and more rigorous plot progression.

\subsubsection{FIERO Frees Stories from Traditional Heroic Tropes} \label{sec:6.1.2}\
Stories from the Control group predominantly mirrored conventional media archetypes, adhering to a standard ``superpowers vs. villains'' framework. For example, C3's characters, though creative, such as \textit{Water Woman, Star Dome, and Xiaoying}, did not break free from traditional sci-fi or superhero genres. \textit{``Water Woman's control over the Ocean Guardian and Star Dome's mastery of light and electricity''} appeared in many traditional sci-fi or superhero works; despite innovations, they largely combined creativity with existing elements. In C3's story, for instance, \textit{``the Dark Witch—a half-human, half-octopus monster hiding in the deepest trench—was a typical vengeance-driven villain and the mastermind behind the conflict. Endowed with the ability to control toxic water and monsters, she sought to eliminate the remaining humans. The three heroes initially acted independently: Water Woman protected a water source base, Star Dome searched the deep sea for the Dark Witch to avenge, and Xiaoying investigated the disaster's truth while searching for her family. The story ultimately unfolded through their collaboration: Water Woman infused pure water into her water strings, Star Dome channeled energy via electricity, and Xiaoying used a prism to convert the electricity into a sacred beam to defeat the witch.''} Despite these creative elements, the characters' motives and emotional conflicts remained simplistic, lacking complex emotional depth and strategic decision-making.

\begin{figure}[t]
    \centering
    \includegraphics[width=1\textwidth]{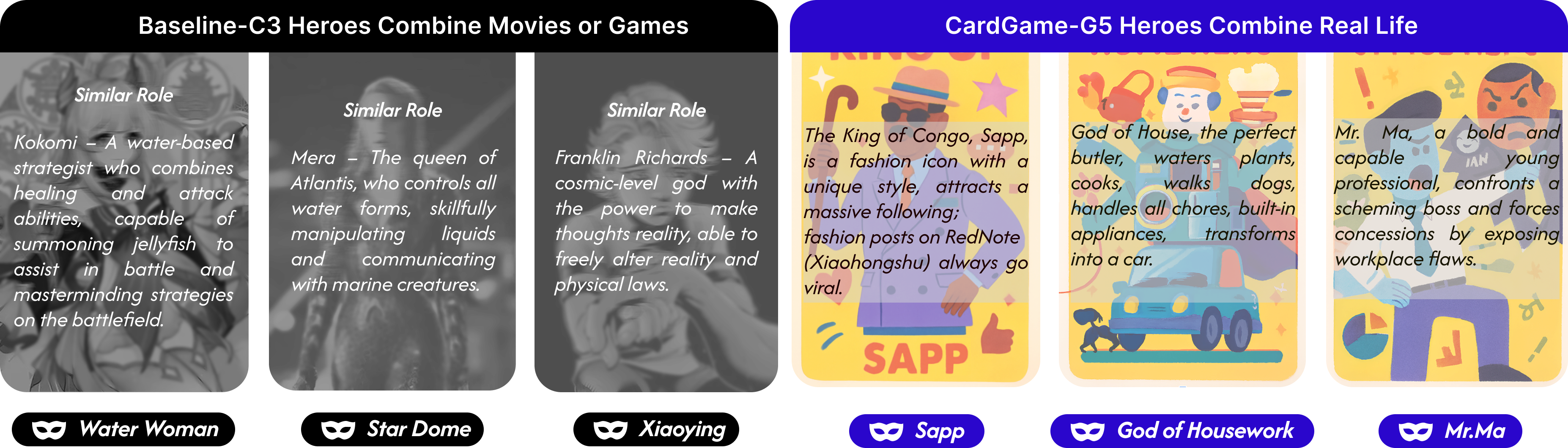}
   \caption{The heroes in the Control group are more closely aligned with existing characters from movies and games, while FIERO's heroes are more connected to real life.}
    \label{fig:story2} 
\end{figure}

FIERO group excelled at integrating daily life elements (e.g., housework, fashion) with superhero abilities, infusing stories with modern life's charm and creativity. For example, G1's \textit{``God of Housework''} combined housework and butler skills with superpowers, breaking free from traditional tropes. Similarly, \textit{``the Fashion God, who used fashion as a core ability to attract fans and wield social influence—was equally inventive''}. The FIERO group's characters transcended standard power-vessel archetypes by integrating mundane, life-oriented traits. In one such story, \textit{``the villain was a businessman who acquired numerous school construction projects by ruthlessly exploiting his subordinates. Envious of the Fashion God Sapp's charm and influence, he believed mastering fashion would allow him to amass wealth and become a social media sensation like Sapp. Driven by jealousy, he plotted to eliminate Sapp, ordering Dr.Ma—a new employee—to sabotage Sapp by seizing their social media accounts and weakening their fan base. Dr.Ma collaborated with the God of Housework, who used their powers to lurk in a restroom and provide precise intelligence. Equipped with AI glasses that could create virtual reality with tangible real-world effects, Dr.Ma used virtual reality manipulation to blow up buildings owned by the businessman. As the structures crumbled, the Fashion God photographed the ruins and posted them on RedNote, where the images went viral, sparking a public opinion storm. This not only boosted the Fashion God's social influence but also exposed the villain's conspiracy.''} By blending traditional superhero tropes with contemporary life elements, the FIERO group innovatively portrayed modern societal roles and plots, as shown in Figure \ref{fig:story2}. Such designs endowed characters with not only superpowers but also deeper social significance and emotional resonance. 

\setlength{\textfloatsep}{4pt}

\added{The narratives generated by the FIERO group reveal the dual impact of the system on users' original creativity. On one hand, FIERO's LLM demonstrated a clear creative amplification. For instance, in group G8, the players initially proposed a relatively sparse original concept: \textit{``Several girls just disappeared from the campus without a sound.''} Assisted by FIERO, this skeletal seed was successfully expanded into a detailed and suspenseful narrative: \textit{``Six consecutive disappearances, each leaving behind a striking bloodstain like a malicious signature carved into the campus cobblestones.''} This textual expansion significantly compensated for non-professional creators' limitations in plot weaving.

On the other hand, this standardized packaging driven by the LLM can inadvertently weaken the unique emotional undertones and narrative vitality present in the raw materials after being refined by the AI. For example, in the story from group G5, the players characterized the protagonist as \textit{``a passionate workplace newcomer who courageously confronts a sinister boss.''} However, AI ``sanitized'' into merely \textit{``a passionate workplace newcomer.''} Similarly, players designed a satirical ending where the boss became an ``ox-horse,'' a Chinese slang for an overworked corporate slave. The LLM lacked meme awareness, so this metaphor was omitted.}

\subsection{\added{Quantitative Findings and Explorations of FIERO}} \label{sec:6.2}

\setlength{\textfloatsep}{4pt}
\begin{figure}[t]
    \centering
    \includegraphics[width=0.7\textwidth]{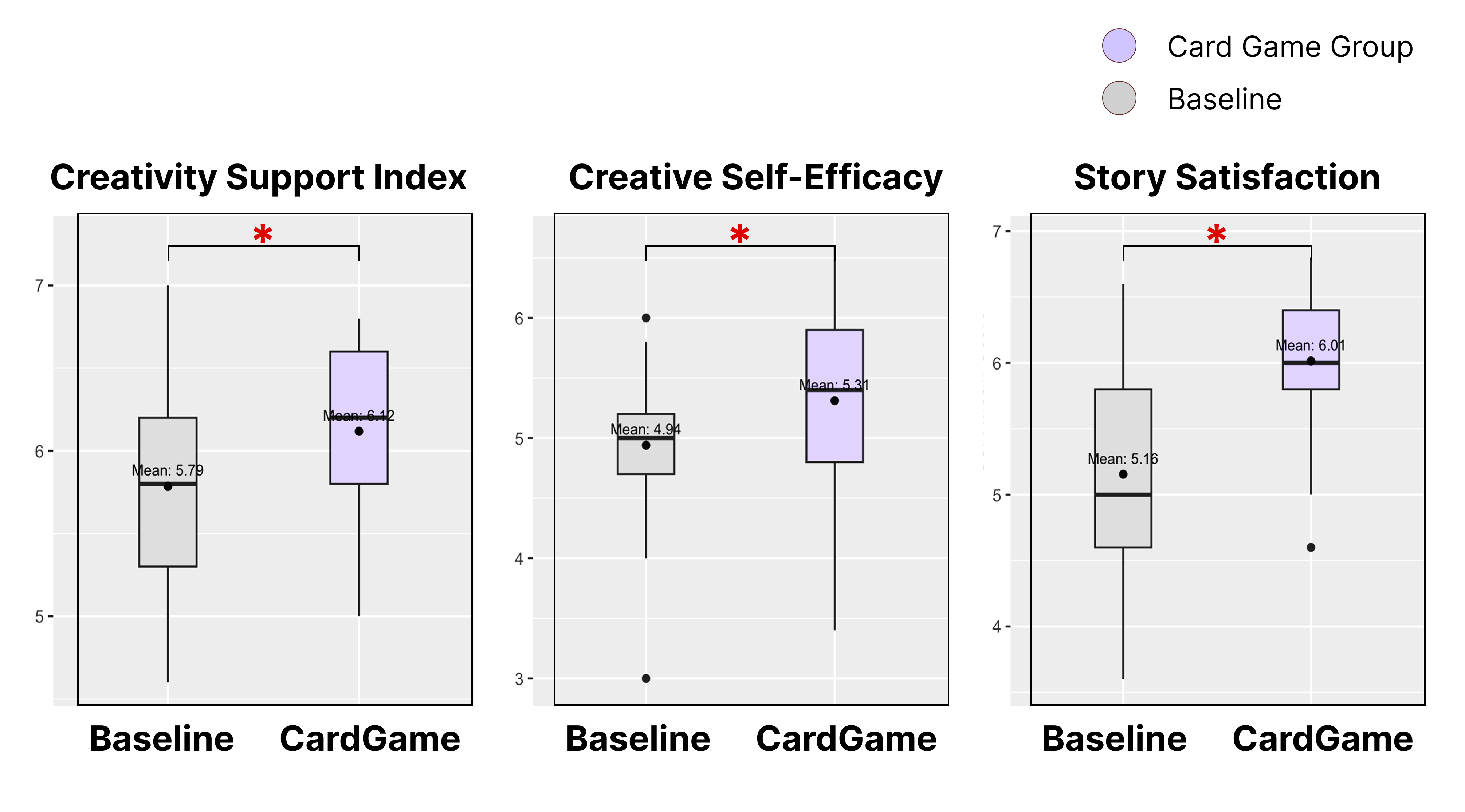}
   \caption{The FIERO group scored significantly higher than the Control group on the Creativity Support Index ($M=6.12$, $SD=0.53$ vs. $M=5.79$, $SD=0.60$, $p=0.032$, $r=-0.276$), Creative Self-Efficacy ($M=5.31$, $SD=0.83$ vs. $M=4.94$, $SD=0.64$, $p=0.047$, $r=-0.256$), and Story Satisfaction Assessment ($M=6.01$, $SD=0.50$ vs. $M=5.16$, $SD=0.90$, $p=0.001$, $r=-0.444$).}
    \label{fig:survey} 
\end{figure}

To evaluate the effectiveness of FIERO in stimulating creative inspiration and facilitating collaborative storytelling within the gaming experience (RQ1), we analyzed survey data using the Mann-Whitney U test and integrated the findings with a qualitative analysis of interview and observational data (see Section \ref{sec:6.3}). \replace{Results from the Mann-Whitney U test indicated that (Figure \ref{fig:survey}), compared to the Control group (N=30), FIERO (N=30) showed significant improvements in terms of Creativity Support Index (CSI) (see Section \ref{sec:6.2}), Creative Self-Efficacy (CSE) (see Section \ref{sec:6.3}), and Story Satisfaction Assessment (SSA) (see Section \ref{sec:6.4}).}{In terms of the total scores across the three core dimensions (CSI, CSE, and SSA), the results from the Mann-Whitney U tests indicated that (Figure \ref{fig:survey}), compared to the Control group (N=30), the FIERO group (N=30) demonstrated significant improvements in Creativity Support Index (CSI), Creative Self-Efficacy (CSE), and Story Satisfaction Assessment (SSA). To gain a deeper understanding, we further conducted an exploratory analysis on individual items within each dimension. Crucially, after applying the Bonferroni correction, the significance threshold was adjusted to $\alpha_{\text{new}} = 0.0033$. Under this stringent criterion, only the ``novelty and originality'' item under the SSA dimension (i.e., SSA2, original $p=0.003$) retained strict statistical significance. Therefore, explorations of other items besides SSA2 should be strictly interpreted as exploratory trends in statistical terms. The specific effect sizes and additional statistical details are provided in Appendix \ref{app:Questionnaire}.}

\subsubsection{\added{Results of the Creativity Support Index (CSI)}}\

\begin{figure}[t]
    \centering
    \includegraphics[width=1\textwidth]{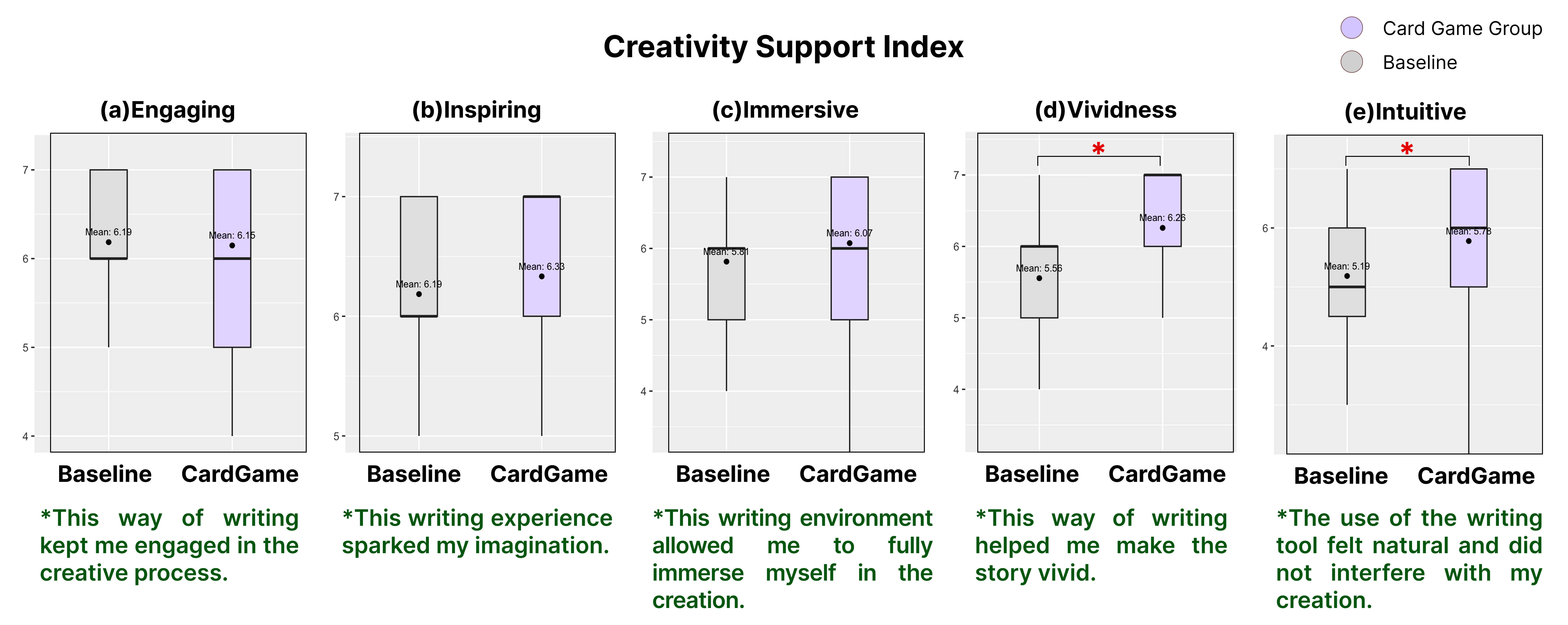}
   \caption{\replace{FIERO showed no significant difference from the Control group in Engaging, Inspiring, and Immersive, but significantly enhanced Vividness ($p=0.011$) and Intuitive ($p=0.046$).}{FIERO performs at a comparable level to the Control group in terms of Engaging, Inspiring, and Immersive, while showing a positive improvement trend in story Vividness and Intuitive stimulation (Exploratory Analysis: Vividness $p = 0.011$, Intuitive $p = 0.046$).}}
    \label{fig:CSI} 
\end{figure}

As shown in Figure \ref{fig:CSI}, the data indicated that, although these differences did not remain significant after the stringent Bonferroni correction, the FIERO group demonstrated an upward trend over the Control group in two metrics: ``Vividness'' ($M_{FIERO} = 6.26, SD = 1.02$ vs. $M_{Control} = 5.56, SD = 1.12$, $p = 0.011$, $r = -0.329$) and ``Intuitive'' ($M_{FIERO} = 5.78, SD = 1.40$ vs. $M_{Control} = 5.19, SD = 1.24$, $p = 0.046$, $r = -0.257$), both achieving medium effect sizes. This positive trend might be attributed to the fact that FIERO provides non-professional creators with richer multimodal sensory stimulation than traditional text-based environments. 

In contrast, the two groups scored remarkably close to each other with no significant differences across three dimensions: ``Engaging'' ($M_{FIERO} = 6.15, SD = 0.95$ vs. $M_{Control} = 6.19, SD = 0.79$, $p = 0.970$, $r = -0.005$), ``Inspiring'' ($M_{FIERO} = 6.33, SD = 0.78$ vs. $M_{Control} = 6.19, SD = 0.68$, $p = 0.363$, $r = -0.117$), and ``Immersive'' ($M_{FIERO} = 6.07, SD = 1.04$ vs. $M_{Control} = 5.81, SD = 0.92$, $p = 0.247$, $r = -0.149$), with corresponding effect sizes being extremely low. This phenomenon likely because face-to-face social collaboration itself offers a high level of engagement.

\setlength{\textfloatsep}{4pt}

\subsubsection{\added{Results of Creative Self-Efficacy (CSE)}}\

\begin{figure}[t]
    \centering
    \includegraphics[width=1\textwidth]{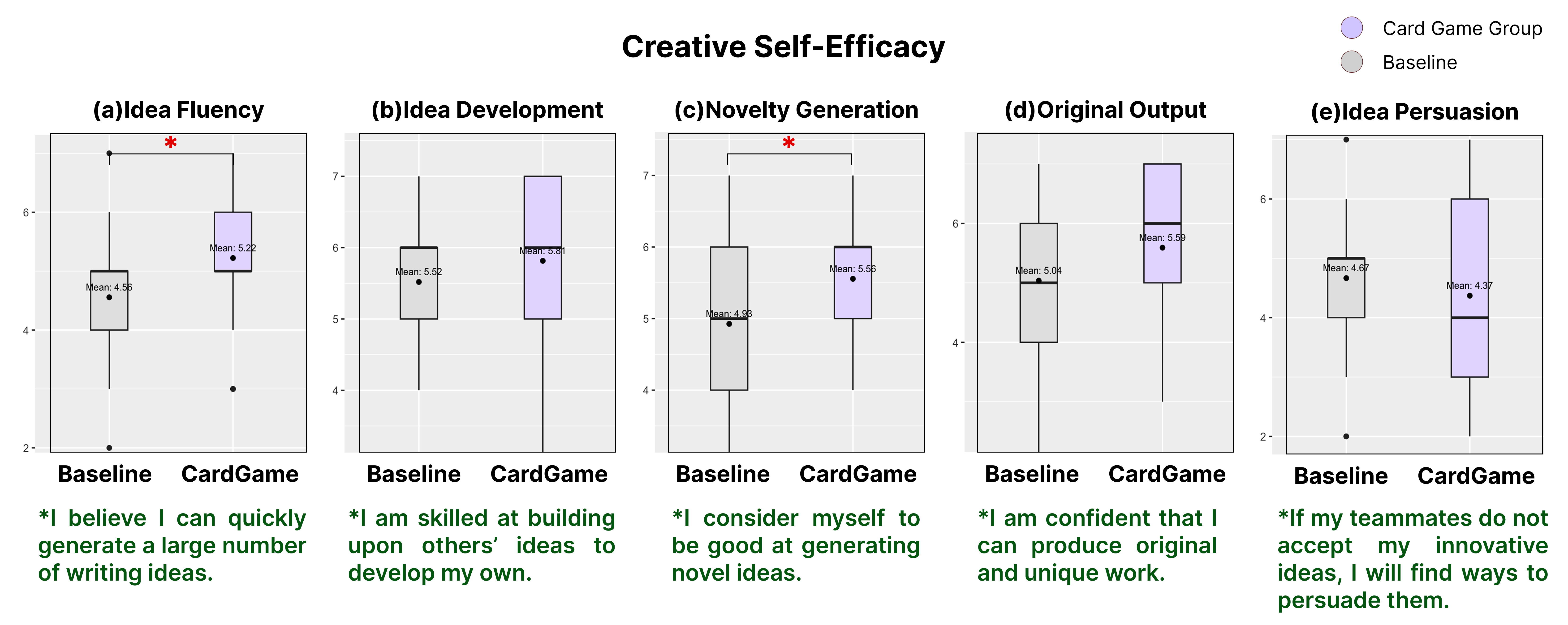}
   \caption{\replace{FIERO showed no significant difference from the baseline group in Idea Development, Original Output, and Idea Persuasion, but significantly enhanced Idea Fluency ($p=0.030$) and Novelty Generation ($p=0.042$).}{FIERO exhibits a similarly high level to the Control group in Idea Development, Original Output, and Idea Persuasion, while demonstrating a positive improvement trend in Idea Fluency and Novelty Generation (Exploratory Analysis: Fluency $p = 0.030$, Novelty $p = 0.042$).}}
    \label{fig:CSE} 
\end{figure}

As illustrated in Figure \ref{fig:CSE}, the data trends revealed that the FIERO group presented a positive inclination of medium intensity in ``Idea Fluency'' ($M_{FIERO} = 5.22, SD = 1.12$ vs. $M_{Control} = 4.56, SD = 1.15$, $p = 0.030$, $r = -0.280$) and ``Novelty Generation'' ($M_{FIERO} = 5.56, SD = 1.09$ vs. $M_{Control} = 4.93, SD = 1.11$, $p = 0.042$, $r = -0.262$). This might be explained by FIERO's implicit card mechanics and its features that assist in linking text, which enhance participants' perceived fluency of ideas, while the randomized guidance inclines them to perceive themselves as capable of conceiving more diverse and novel concepts.

However, the two groups did not exhibit any distinct gaps in the dimensions of ``Idea Development'' ($M_{FIERO} = 5.81, SD = 1.33$ vs. $M_{Control} = 5.52, SD = 1.22$, $p = 0.253$, $r = -0.148$), ``Original Output'' ($M_{FIERO} = 5.59, SD = 1.22$ vs. $M_{Control} = 5.04, SD = 1.19$, $p = 0.105$, $r = -0.209$), and ``Idea Persuasion'' ($M_{FIERO} = 4.37, SD = 1.52$ vs. $M_{Control} = 4.67, SD = 1.24$, $p = 0.431$, $r = -0.102$). This parity could stem from the fact that during group co-creation, players can heavily rely on mutual interpersonal discussions to refine story details, persuade teammates, and advance the script through shared brainstorming, a core social foundation that remains unaltered by the introduction of the tool.

\setlength{\textfloatsep}{4pt}

\setlength{\textfloatsep}{4pt}
\begin{figure}[t]
    \centering
    \includegraphics[width=1\textwidth]{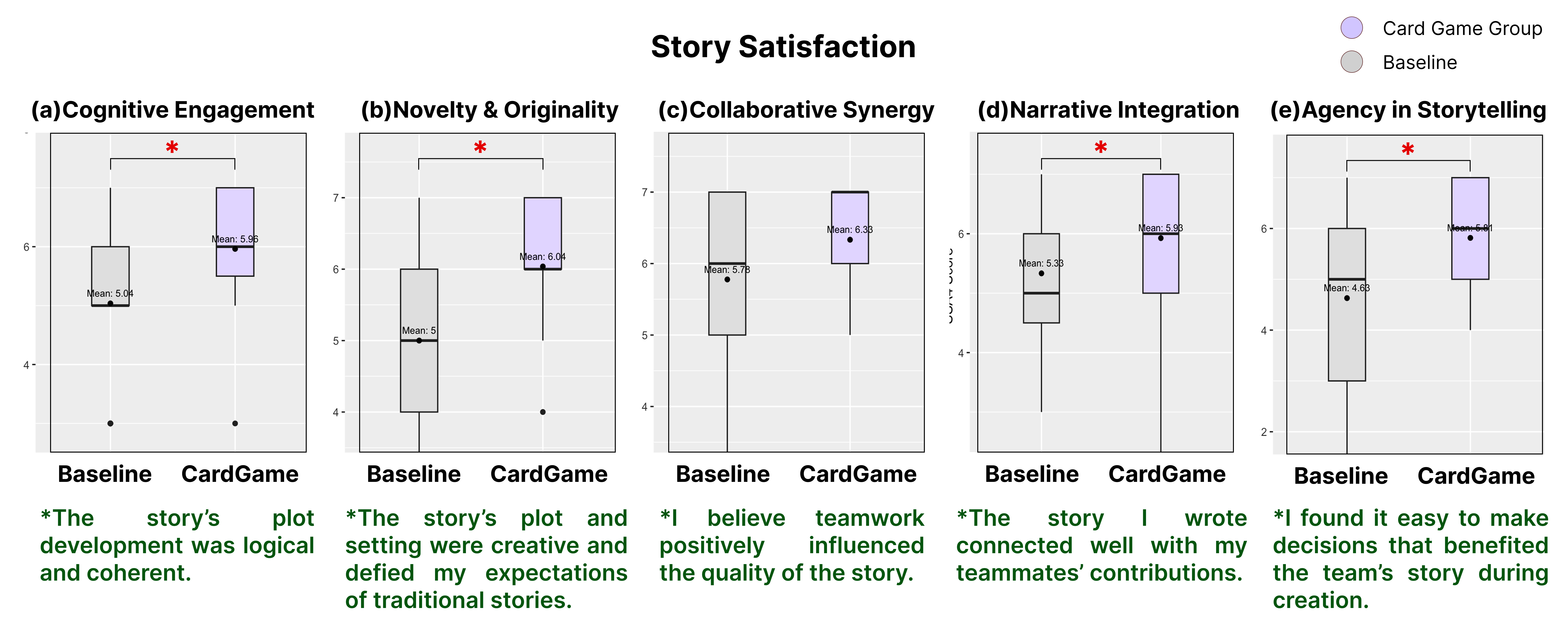}
   \caption{\replace{FIERO showed no significant difference from the Control group in Collaborative Synergy, but significantly enhanced story Cognitive Engagement ($p=0.010$), Novelty \& Originality ($p=0.003$), Narrative Integration ($p=0.029$), and Agency in Storytelling ($p=0.013$).}{FIERO performs at a comparable level to the Control group in collaborative synergy; following the Bonferroni correction, FIERO significantly enhances story Novelty \& Originality ($p = 0.003$), while exhibiting a positive improvement trend in Cognitive Engagement, Narrative Integration, and Agency in Storytelling (Exploratory Analysis: Engagement $p = 0.010$, Integration $p = 0.029$, Agency $p = 0.013$).}}
    \label{fig:SSA} 
\end{figure}

\subsubsection{\added{Results of Story Satisfaction Assessment (SSA)}}\

As shown in Figure \ref{fig:SSA}, the FIERO group scored significantly higher than the Control group across four dimensions: Cognitive Engagement ($M_{FIERO} = 5.96, SD = 0.94$ vs. $M_{Control} = 5.04, SD = 1.43$, $p = 0.010$, $r = -0.333$), Narrative Integration ($M_{FIERO} = 5.93, SD = 0.96$ vs. $M_{Control} = 5.33, SD = 1.11$, $p = 0.029$, $r = -0.282$), and Agency in Storytelling ($M_{FIERO} = 5.81, SD = 1.08$ vs. $M_{Control} = 4.63, SD = 1.76$, $p = 0.013$, $r = -0.321$). This superior performance suggested that without depriving creators of their narrative ownership and agency, FIERO appeared more capable of transforming fragmented creative ideas into logically coherent narrative outcomes. Notably, the ``Novelty \& Originality'' item ($M_{FIERO} = 6.04, SD = 1.02$ vs. $M_{Control} = 5.00, SD = 1.30$, $p = 0.003$, $r = -0.388$) retained strict statistical significance even after undergoing the rigorous Bonferroni multiple comparison correction (threshold adjusted to $\alpha_{\text{new}} = 0.0033$), highlighting a potential advantage of FIERO in stimulating intrinsic innovation and breaking narrative expectations.

Regarding the Collaborative Synergy dimension, no significant difference was observed between the two groups ($M_{FIERO} = 6.33, SD = 1.02$ vs. $M_{Control} = 5.78, SD = 1.28$, $p = 0.74$, $r = -0.231$). This confirms that even when the interaction media and visual presentations change, the innate human social tacit understanding and team bonding under the offline physical collaboration modality remain at a consistently high level.

\subsection{\added{Qualitative Findings on Collaborative Creative Writing}} \label{sec:6.3}

\subsubsection{\added{Perceptions of Writing Modalities}}

\added{Participants generally reflected that FIERO's ``cards help stimulate imagination,'' thereby disrupting their conventional mindsets and expanding their creative boundaries.} The randomness of the cards was a highlight in the creative process, breaking fixed ways of thinking and prompting participants to produce more absurd and unexpected content. When certain card combinations appeared, creativity often surpassed traditional frameworks, leading to unpredictable story developments. While this randomness brought freshness, it sometimes also caused confusion. As player B from G5 said: \textit{``The randomness of the cards made our story more absurd, sometimes chaotic, but also interesting.''} The cards' use prevented the creation from being limited to participants' habitual character setups. Some expressed that without the cards' guidance, they might have simply recreated characters from movies or animations they had seen. The unique characters and plot settings provided by the cards helped them break out of their comfort zones and expand their imaginative boundaries. As player B from G6 said: \textit{``Without the cards, I might unconsciously recreate movie characters, but the cards pushed me out of that box to try different setups.''} Player C from G1 also clearly stated: \textit{``Without these cards, I wouldn't have come up with these ideas. Without the cards, I couldn't think of these points from scratch.''} This showed that the specific information provided by the cards (such as character traits and ability setups) gave the creation a clear starting point. For example, when the physical card showed \textit{``an extra leg,''} player C associated it with \textit{``possible new abilities and skills.''} This association triggered by the physical attributes of the card directly propelled the storyline's development.

\added{When collaborative decision-making reached an impasse, teams in both groups expressed that ``someone needs to lead and drive the process,'' a role that FIERO appeared to fulfill more effectively in some cases.} During gameplay, three-choice decisions on storylines often left teams in silence and hesitation. FIERO played the role of an ``icebreaker'' here: its generated suggestions were often \replace{highly persuasive}{perceived as persuasive}, helping teams quickly reach a unified decision and advance the story. However, this effectiveness was moderated by team agency: groups that were more talkative and creative tended to rely on their own discussions rather than adopting FIERO's suggestions. A similar pattern was observed in the Control group: once someone proposed a plan, other members quickly agreed. Overall, neither group exhibited a sustained ``persuasion game''; instead, both relied more on the formation of rapid consensus. 

\subsubsection{\added{Interactive Behaviors in the Creative Process}}

\begin{figure}[t]
    \centering
    \includegraphics[width=1\textwidth]{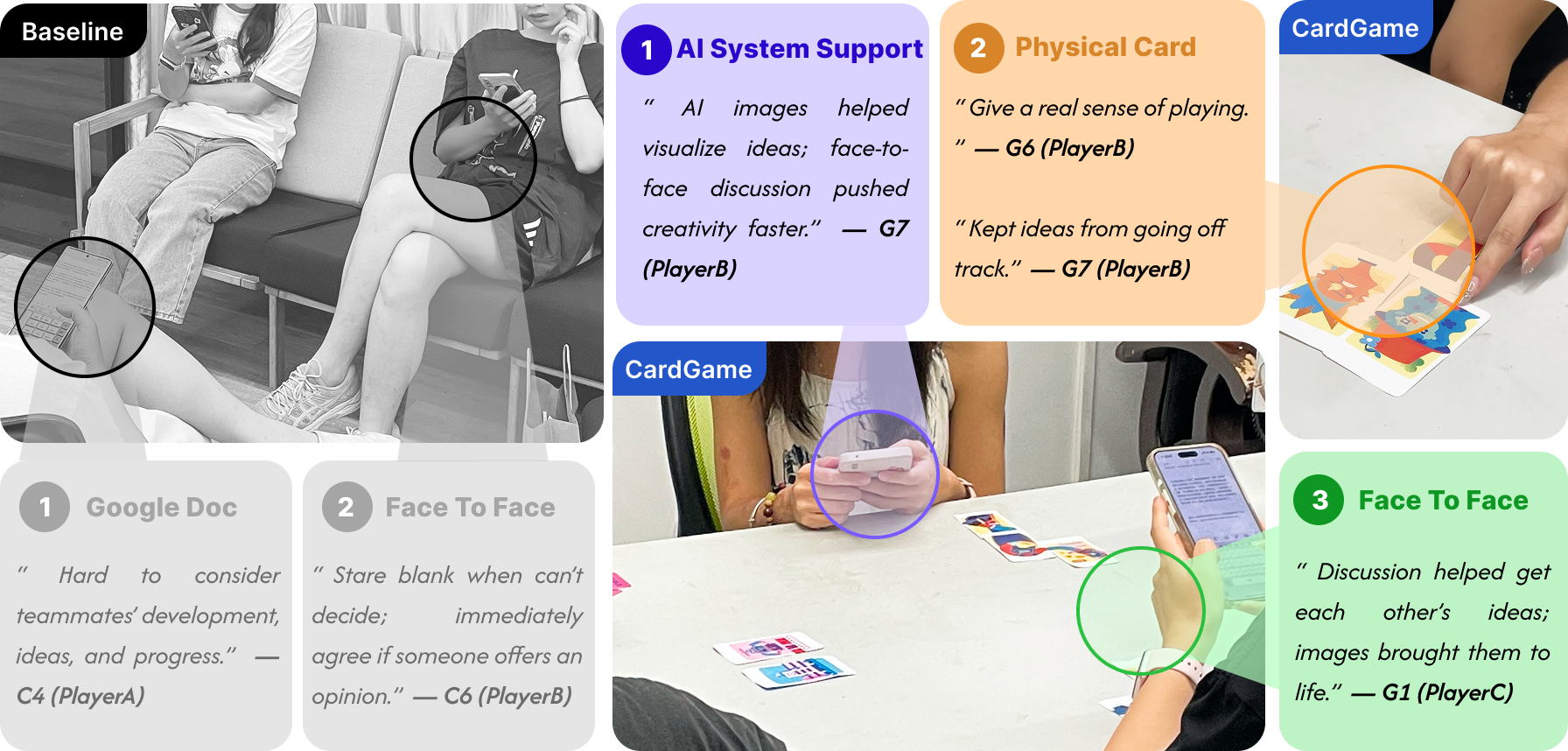}
   \caption{Intuitive Interaction Sparks Creativity and Immersion: A comparison between the FIERO group and the Baseline group under diverse creativity stimulation conditions.}
    \label{findings1} 
\end{figure}

In the FIERO group, the combination of cards, AI-generated images, and peer interaction offered a diverse multi-sensory experience, as shown in Figure \ref{findings1}. The cards acted as a creative framework, helping teams quickly focus and avoid excessive divergence. The AI-generated images transformed abstract ideas into concrete visual elements, enabling team members to more intuitively understand characters and scenes. Interactions among teammates, through \added{``eye contact and facial expressions'' alongside} verbal communication, strengthened emotional bonds and sparked creative collisions, driving the creative process forward. Many participants noted that face-to-face communication, through the interplay of multiple elements, stimulated and integrated ideas more quickly. \added{Crucially, although participants in both groups could access each other's content online, those in the co-located setting were far more prone to spontaneously engage in the behavior of ``looking at each other's phones.''} Each of these interaction points increased the potential for innovation and heightened the sense of excitement in the experience. For example, Player B from G7 remarked: \textit{``The cards helped us avoid ideas straying too far, the AI images helped us materialize our concepts, and face-to-face discussions accelerated our creative process.''} Similarly, Player A from G6 commented: \textit{``Face-to-face discussions made it easier for us to understand each other's ideas, while the images helped us make them concrete rather than leaving them as abstract concepts.''}

 \setlength{\textfloatsep}{4pt}

\added{Furthermore, concrete hands-on practices such as ``discussing while pointing at the cards'' elevated the cards from mere prompts to shared visual anchors.} By pointing to and gesturing at the cards to describe their ideas, many participants felt a stronger sense of interaction. This hands-on engagement elevated the cards beyond mere prompts—they became shared visual anchors, making it easier to communicate ideas, reach consensus, and refine them in real time. The combination of tactile and visual experiences not only boosted creativity but also deepened the sense of co-creation within the team. Player C from G1 added: \textit{``I really liked the physical cards; they gave me a sense of truly participating.''} This suggests that interaction with physical artifacts strengthened the connection between creation and tools, thereby reconfiguring the behavioral patterns of collaborative interaction.

Since players in both groups could observe each other's writing processes, many were able to tacitly develop similar themes. For example, C4 collectively set their superhero story against the backdrop of an alien invasion in their local city. However, this process also revealed that one's own ideas could be influenced by collaborative writing. Participants across both groups generally agreed that face-to-face communication and real-time discussions allowed them to quickly share ideas, promptly correct misunderstandings, and strengthen mutual understanding of the story. Especially when defining characters and plots, teammates' suggestions and perspectives often brought unexpected inspiration, injecting new vitality and emotion into the creative process. As Player C from G1 stated, \textit{``One person can only come up with one type of story, but with one more person, there's one more possibility.''}

\subsubsection{\added{Reflections on Creative Outcomes}}

\added{For non-professional creators, ``inspiration is rooted in life'' is a commonly observed phenomenon, where the initial ideas proposed by participants were often drawn directly from their respective real-life experiences, common sense, or existing movie plots. In the Control group (without FIERO), collisions of personal inspirations led to a ``difficulty in merging worldviews.'' FIERO showed a clear ability to ``refine structure'' and ``supplement logic'', turning fragmented thoughts into a complete narrative.} AI helped organize and connect creative ideas, transforming fragments into a coherent story. Player A from G1 mentioned: \textit{``The final story outline was basically based on the AI's modification... otherwise, the initial ideas would have been scattered, and it helped organize them into a story with a beginning, climax, and conclusion.''} Player C from G1 added: \textit{``When I write by myself, I definitely miss a lot of things, like the climax and conclusion... AI helped me complete it,''} indicating that AI solved the structural flaws caused by limited time (\textit{``Five minutes of brainstorming is not enough''}) or fragmented thinking. AI also provided creative satisfaction, especially for those lacking initial inspiration. For participants who were unsure, the ``satisfaction'' was particularly evident through AI integration. For ``directionless'' creation (as described by player A from G2), AI helped fill in logic and expand details. Player A mentioned: \textit{``My ideas were scattered, lacking logic... AI helped me supplement and expand my ideas.''} For example, it transformed the vague idea of ``a hero defeating someone'' into a logically structured narrative. FIERO was especially \replace{crucial}{helpful} for uninspired players, supplementing logic and expanding details to overcome blocks. As Player C from G1 stated, \textit{``I really had no ideas at first, so I was surprised when a complete story finally took shape—it exceeded my expectations.''} FIERO helped them complete the creation ``from a different perspective.''

\added{This capability of FIERO to ``refine structure'' and ``supplement logic'' is further demonstrated by the fact that,} despite the highly random nature of the cards, they are part of a design system, with each card having certain potential connections and logic with others. When using the cards, participants often noticed these hidden connections, which helped make their stories more cohesive and logical during the creative process. The cards were not only a source of inspiration but also an invisible structure and framework for the story. As player A from G1 said: \textit{``The cards seem random, but the hidden connections and logic made our creative process more organized.''}  In this game, AI and the cards provided a structured creation process, which was crucial for effectively advancing creativity. Cards helped focus theme and plot, allowing quick story direction decisions in limited time. AI organized fragmented thoughts into coherent stories, ensuring character and plot consistency. As player B from G6 said: \textit{``The cards helped us quickly focus within a framework, and AI then connected our fragmented ideas into a coherent story.''} \replace{This explains}{This may help explain} why FIERO helps players connect and make decisions beneficial to the team's story. Unlike the Control group (adapting from films/TV), FIERO guided players to break away from familiar settings and create original roles and plots.

However, it is noteworthy that FIERO demonstrated limitations in certain scenarios, such as avoiding content generation and assistance with more bold or radical topics. For instance, Player B from G8 mentioned wanting to create a serial killer-like character but noted that \textit{``the AI couldn't help with overly radical elements.''} Some players also pointed out limitations in FIERO's visual stimulation. As Player B from G5 put it, \textit{``The AI only converts our text descriptions into images, but it doesn't drive the plot forward—we still rely mostly on our own imagination,''} indicating that FIERO's assistance is limited in some cases. Furthermore, participants also perceived gender bias in AI visual generation. For example, when players created gender-neutral superheroes, the AI consistently depicted them as male (e.g., magicians, heroes with stacked abilities, new employees, leaders, and farmers).

\subsubsection{\added{Emotional Experiences in the Creative Process}}

\added{For both groups, ``creative relaxation'' and ``absurdity is fun'' intertwined to form the participants' emotional experience.} Participants noted that they could conceive absurd, whimsical story elements, and the creative process itself became a form of relaxation and brainstorming. This relaxed, free-form approach allowed them to temporarily escape real-world pressures and enjoy the pleasure of creation. No matter how absurd the story became, they still found joy in the process and stumbled upon unexpected inspiration. For instance, Player B from G8 stated, \textit{``I usually do mechanical work and feel a bit numbed, so this creative experience was really great.''} Player B from G5 \textit{also felt that the absurd parts of the story were the most entertaining}. \added{Meanwhile, collaboration within this co-located environment contributed to an ``enhanced social attribute,'' as} Player B from G7 commented, \textit{``Online communication is tricky and can feel awkward, but face-to-face interactions let us communicate more naturally.''} 

\begin{figure}[t]
    \centering
    \includegraphics[width=1\textwidth]{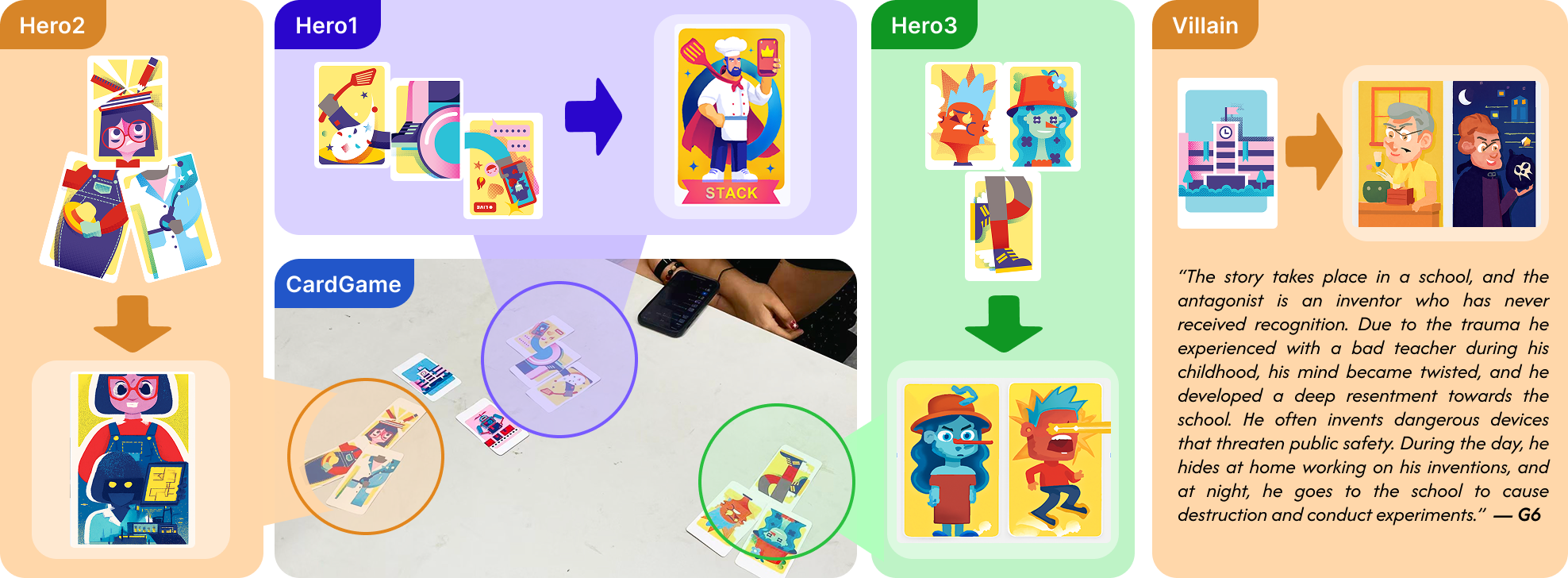}
   \caption{FIERO enhances the Vividness of the story through cards and AI-generated content. \added{The three drawn hero cards combined with player text were utilized to generate a new character image card for each player. Concurrently, the location card combined with player text was used to generate new villain image cards, from which one final card was selected.}}
    \label{findings2} 
\end{figure}

\added{Beyond this, driven by the card images generated by FIERO, the FIERO group additionally reported an emotional feeling of ``sense of immersion.''} This sense of immersion may be attributed to AI-generated images, which played a key role in transforming abstract concepts into concrete visual forms, as shown in Figure \ref{findings2}.  They helped participants better understand characters and scenes, strengthening the visualization of their creations. While these images did not introduce groundbreaking ideas for the plot, they enriched emotional expression and character design. For example, Player A from G7 noted that the combination of cards and AI images was like a \textit{ ``storyboard or mood board,''} filling their mind with ideas. Many players expressed a desire to continue writing\textit{ ``if time hadn't been limited.''}

\setlength{\textfloatsep}{4pt}

Participants generally agreed that AI images supported story construction and character portrayal, particularly in visualizing plot elements. As Player C from G6 noted, \textit{``AI-generated images directly showed what the characters looked like, helping us better understand and build the story.''} However, some players found it hard to concentrate. For example, Player A from G6 said, \textit{``The game was too short; I hadn't fully immersed myself, so the experience felt shallow.''}

\subsection{Game-Based Collaboration Enhances Story Outcomes} \label{sec:6.4}

To address RQ2, we conducted large-scale pairwise evaluations between the \textbf{FIERO condition} and the \textbf{control condition} using three independent LLM judges: Claude Opus~4.5, Gemini~3~Pro, and GPT-5.2. Each comparison followed a strict AB/BA cross-comparison protocol to control for positional bias, where inconsistent outcomes between forward (A$\rightarrow$B) and reverse (B$\rightarrow$A) comparisons were labeled as draws.

\subsubsection{Overall Comparison Outcomes}

\begin{figure}[!h]
\centering
\includegraphics[width=\linewidth]{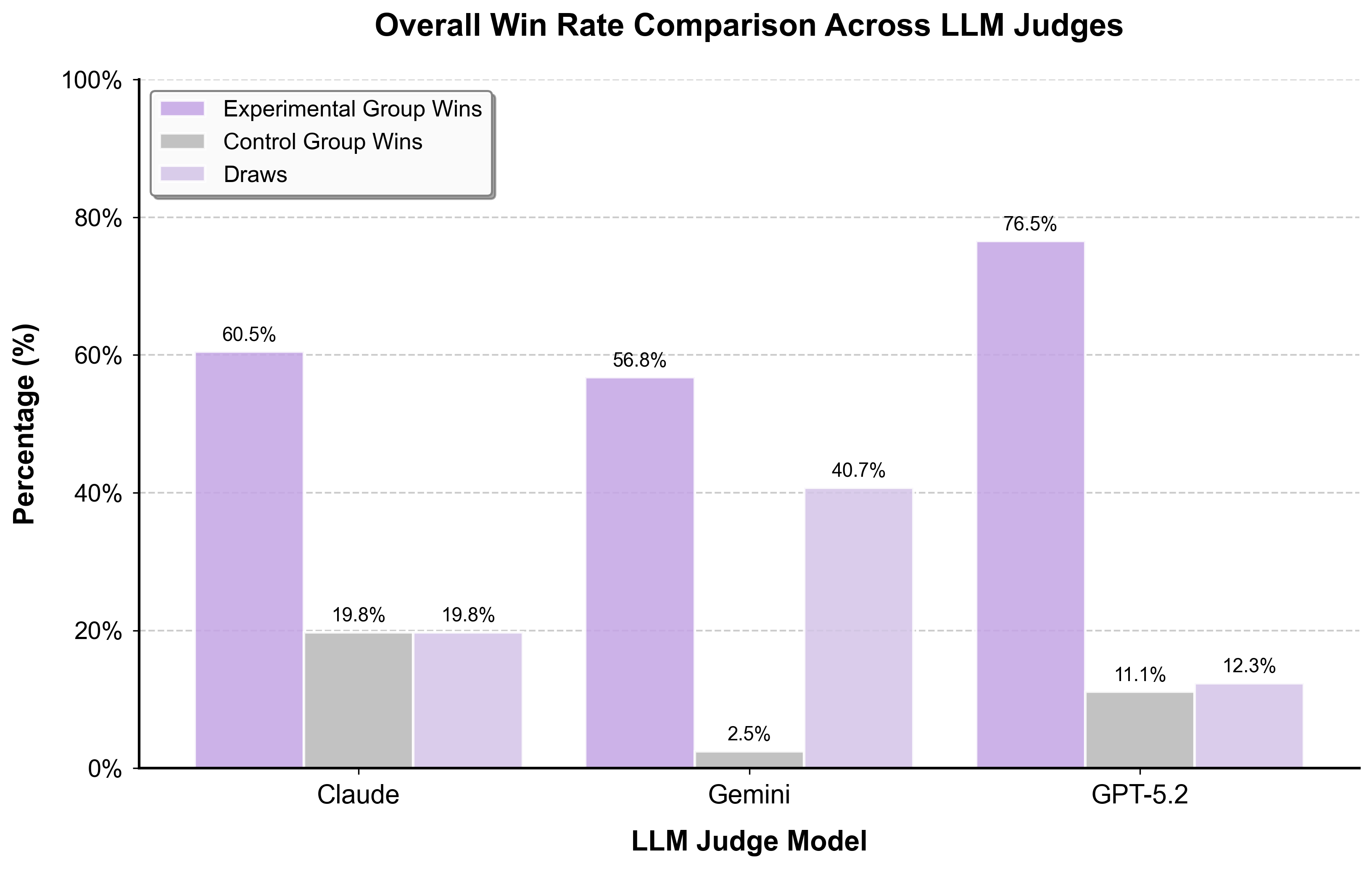}
\caption{\added{Detailed win rate comparisons across all seven dimensions for each LLM judge.}}
\label{fig:overall}
\end{figure}

Across all judges, the FIERO condition consistently outperformed the Control condition (Figure \ref{fig:overall}). Specifically, the FIERO condition achieved win rates of 60.5\% (49/81) with Claude, 56.8\% (46/81) with Gemini, and 76.5\% (62/81) with GPT-5.2. In contrast, the control condition showed substantially lower win rates across all models. Draws accounted for between 12.3\% and 40.7\% of comparisons.


While absolute win rates varied, all three judges exhibited the same overall pattern favoring the FIERO condition, suggesting that the observed differences reflect genuine characteristics of the stories rather than idiosyncrasies of individual evaluators. Chi-squared tests confirmed that the observed differences between conditions were statistically significant for all three judges (Claude: $\chi^2 = 16.00$, $p < .001$; Gemini: $\chi^2 = 38.35$, $p < .001$; GPT-5.2: $\chi^2 = 39.56$, $p < .001$). These results indicate that the patterns described above are unlikely to be attributable to chance.

\subsubsection{Dimension-Level Results}

\begin{figure}[!h]
\centering
\includegraphics[width=\linewidth]{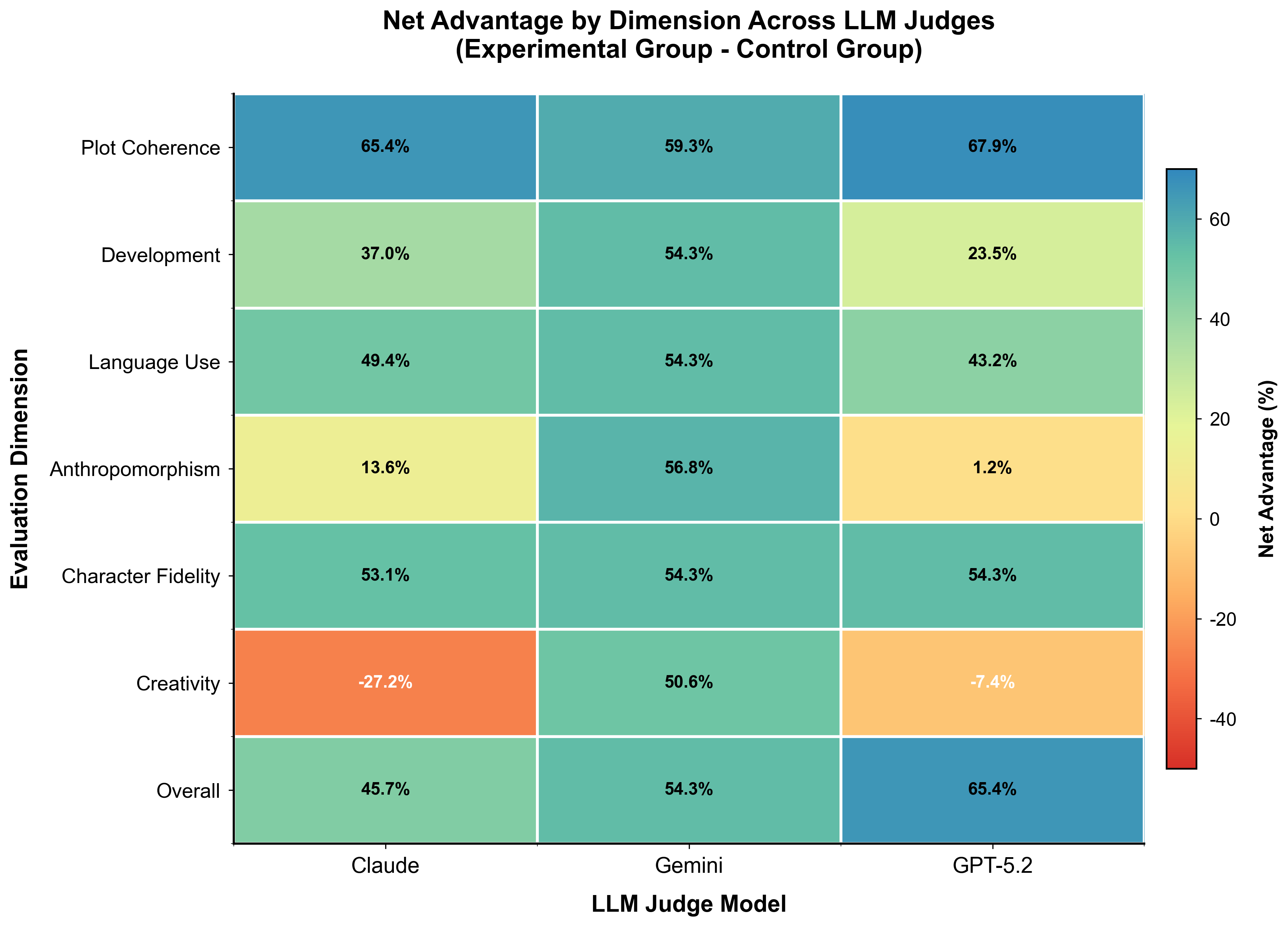}
\caption{\added{Net Advantage by Dimension Across LLM Judges. Percentage net advantage of the FIERO condition over the control condition, evaluated across seven dimensions by three LLM judges. Color intensity represents magnitude; green indicates FIERO advantage, red indicates control advantage. The FIERO condition showed consistent positive advantages for Plot, Character Fidelity, Language Use, and Overall Preference. Creativity exhibited the greatest inter-judge variability, with Gemini favoring FIERO (+50.6\%) while Claude and GPT-5.2 favored the control condition (-27.2\% and -7.4\%).}}
\label{fig:dimension}
\end{figure}

Stories generated under the FIERO condition and the control condition were evaluated across seven dimensions: Plot, Development, Language Use, Anthropomorphism, Character Fidelity, Creativity, and Overall Preference. Each dimension was assessed independently, with outcomes recorded as win, loss, or draw. Figure \ref{fig:dimension} presents the complete win-rate comparisons across all dimensions.

\paragraph{\textbf{Plot Coherence}}
Across all judges, the FIERO condition demonstrated a clear advantage in Plot. Net advantages ranged from 59.3\% to 67.9\% \added{(representing the percentage margins by which FIERO outperformed the Control condition)}. This result indicates that FIERO-generated stories are more likely to feature clear narrative structures with logically coherent plot progression, well-developed conflicts and resolutions, and smooth narrative transitions. For instance, one FIERO-group story exhibited a complete narrative arc: ``setup'' (introducing the three heroes and antagonist), ``development'' (Dr.Ma tasked with confronting the Fashion God), ``twist'' (Dr.Ma defecting and launching a counterattack with allies), and ``resolution'' (the antagonist's conspiracy exposed, heroes victorious). In contrast, multiple control-group stories suffered from structural ambiguity, with some endings stopping at the beginning of a battle scene rather than providing resolution, or terminating after merely establishing background settings.

\paragraph{\textbf{Character Fidelity}}
Consistent improvements were observed for Character Fidelity under the FIERO condition, with net advantages exceeding 50\% across all three judges. This suggests that characters in FIERO-generated stories are more likely to maintain internal consistency between behavior and personality, with decisions and actions exhibiting greater credibility. The Control group exhibited multiple instances of ``character collapse'': some characters' core functions were erased during story climaxes, reducing their presence to zero; others involved antagonists with confused identities or contradictory life-death states across different segments, rendering the worldview self-contradictory.

\paragraph{\textbf{Language Use}}
The FIERO condition also showed strong performance in Language Use, with net advantages between 43.2\% and 54.3\%. Stories generated under this framework tended to employ richer and more precise vocabulary with greater syntactic variety, achieving clear expression with literary appeal. For example, FIERO-group stories featured concrete expressions such as ``eternal night,'' ``primordial darkness,'' and ``fashion sensibility.'' In contrast, control-group stories frequently relied on hollow clichés like ``mysterious,'' ``super invincible,'' and ``badly battered,'' along with run-on narratives characterized by oral connectives such as ``and... meanwhile... also...''

\paragraph{\textbf{Development}}
For Development, including world-building and story progression, results were more moderate but still favored the FIERO condition. Net advantages ranged from 23.5\% to 54.3\%, with Gemini exhibiting the strongest preference. FIERO-generated stories more frequently demonstrated deeper world-building, richer character backgrounds, more nuanced environmental descriptions, and more layered story progression. For instance, FIERO-group descriptions of the ``workplace world'' included concrete rules such as ``ineffective management,'' ``employees overwhelmed with misery,'' and ``capable newcomer vs. unreliable leadership.'' The Control group, however, exhibited shallow world-building with vague descriptions such as ``The world was left in tatters by the destruction of a dark witch, who released sources of pollution all over the world. Aquagirl's family was killed by the pollution,'' along with absent character backgrounds and rushed plot progression.

\paragraph{\textbf{Anthropomorphism}}
Results for Anthropomorphism were mixed. Claude and GPT-5.2 showed relatively small differences between the FIERO and control conditions (net advantages of 1.2–13.6\%), while Gemini strongly favored the FIERO condition (net advantage of 56.8\%). FIERO-group stories featured psychologically nuanced descriptions such as ``working hard to hawk his wares,'' ``hedgehog-like,'' and ``sharp and worldly-wise,'' while control-group characters appeared relatively flat, ``without emotions,'' ``desires,'' or ``inner struggles.'' However, the overall gap in personification performance between the two conditions remained unstable.

\paragraph{\textbf{Creativity}}
Creativity exhibited the greatest variability across judges. Gemini favored the FIERO condition (net advantage of 50.6\%), whereas Claude and GPT-5.2 favored the control condition (net advantages ranging from -27.2\% to -7.4\%). FIERO-group stories featured original concepts such as ``permanently using a wheelchair because her lower limbs store power for her ultimate move, while misleading enemies into letting down their guard'' and ``hair that can grow beautiful flowers for camouflage.'' Control-group stories more often adhered to traditional narrative frameworks, but drew inspiration from folk archetypes, gaming culture, and popular media to create imaginative characters such as ``Thousand-Mile Ear'', ``Pigman'', and an ``Ender Man capable of stealing blocks''. Although these characters were rooted in recognizable cultural references, they may still have been perceived as innovative by the judges.

\paragraph{\textbf{Overall}}
Despite dimensional variations, Overall Preference consistently favored the FIERO condition across all judges (45.7–65.4\% net advantage), indicating that stories produced through game-based collaboration more compelling.

\section{Discussion}\label{sec:Discussion}
FIERO introduces a hybrid framework that integrates tabletop card gameplay with an AI-powered digital system for collaborative storytelling. The framework is designed to address key challenges in collaborative narratives, such as narrative fragmentation and uneven participation, while balancing structured guidance with creative spontaneity. By synthesizing insights from prior research on human-AI collaboration, creative writing, and multimodal interaction, this work contributes to understanding how physical-digital hybrid tools can enhance collective storytelling.

\subsection{Bridging Physical and Digital: Multimodal Collaboration}

Cognitive load theory suggests that when information is presented in a multimodal way, it can reduce the load on working memory, allowing thoughts to focus more on idea generation without being distracted by irrelevant details \cite{sweller2011cognitive}. \replace{FIERO's core innovation}{A key feature of FIERO} lies in its integration of tangible card-based mechanics with an AI-powered digital system, a design choice rooted in evidence that multimodal interaction strengthens creative engagement \cite{Chung2022TaleBrush,qin2024charactermeet}. \added{On a cognitive level, this physical-digital duality triggers a cross-sensory resonance between physical symbols and digital content \cite{Waterworth1997Creativity}.}

Existing research indicates that technology has transformed our physical interactions into infinitely expandable and flexible digital interactions. We can use AI to create an unlimited number of stories. However, these digital experiences lack physical elements like flipping through pages and social interaction \cite{Gupta2020Digitally}. Physical cards provide a low-friction, socially intuitive interface for idea exchange—aligning with findings from Toyteller that tangible manipulation (e.g., moving character symbols) reduces creative anxiety and enables expression of ``vague intentions'' difficult to verbalize \cite{chung2025toyteller}. As participants in this study stated, the presence of the cards enhanced the creative experience (G1 Player C) and helped them better communicate and spark ideas with teammates (G6 Player A). \added{As demonstrated by Augisoro \cite{braithwaite2025augisoro}, an augmented traditional tabletop game system, the natural tactile affordances and spatial openness of physical mediums, including tokens or cards, provide the most intuitive embodied negotiation interface for multi-user collective meaning-making.} Meanwhile, the role of the digital system in collaborative writing, visualizing scenes, resolving conflicts, and weaving fragmented ideas into coherent arcs addresses a longstanding gap in traditional collaborative tools, which often prioritize text sharing over narrative cohesion \cite{gero2023social,chung2025toyteller}. \replace{This study bridges this gap through the combination of physical and digital mechanisms—although collaborative writing can rely on large amounts of text, it often faces challenges such as communication breakdowns, fragmented creativity, and lack of coherence \cite{Park2023why}.}{Although group collaborative writing inherently faces challenges such as communication breakdowns and fragmented creativity \cite{Park2023why}, this study addresses this gap through the integration of physical and digital mechanisms, contributing to the overall coherence of the finalized story while supporting high-frequency, face-to-face negotiation among co-located users.}

\replace{The duality mirrors previous works with 3D avatars and voice interaction, where multimodal stimuli (visual, auditory, textual) collectively enhance immersion \cite{qin2024charactermeet}. In FIERO, the visuals of the cards together with digital system generated visuals create a cross-sensory resonance \cite{Waterworth1997Creativity}, triggering richer associative thinking among participants. For example, a character card paired with digital system rendered heroes which is not perfectly fited to participants' imagination prompted groups to invent more nuanced plot twists than text-only prompts (G1 Player A, although noting that the AI-rendered character had an extra leg compared to his imagined one, thought this change added a new characteristic to the hero). This phenomenon is similar to the ``mild place illusion'' found in research, which strengthens creative immersion \cite{Goncalves2018Mild}.}{However, while participants in this study did not report a significant workflow burden from multimodal collaboration, viewing it instead primarily as a source of novelty, multimodal synergy is not without its costs \cite{oviatt2004we}. Research suggests that although multimodal input can alleviate information pressure from a single channel and assist in attention allocation and spatial understanding, it can conversely increase users' switching and coordination burdens if poorly designed \cite{lazaro2024mind}. Recent evidence in multimodal creative writing indicates that while expanding interaction modalities enhances users' expressiveness and immersion, it inherently imposes a heavier cognitive demand \cite{fu2026vistoria}. As collaborative writing tasks escalate or experimental sessions extend over time, players operating systems may gradually perceive the cognitive costs brought by frequent switching between physical cards and digital UI inputs. When interactions cycle at high frequencies between tangible objects and digital interfaces, the brain requires continuous attention resetting and working memory clearing \cite{radvansky2001working}, which to some extent creates fragmented disruptions. Consequently, in more complex task streams, the inherent advantages of multimodal interaction may be progressively eroded, eventually evolving into an additional mental workload \cite{zimmerer2022reducing}. This potential cognitive boundary warrants deeper empirical investigation in future research.}

\subsection{Iterative Creativity and Life-like Narrative}

The design of FIERO adopts an iterative process, where group members generate ideas through card games as the process progresses, and can use the digital system to integrate and refine these ideas. Each stage requires discussion to connect the contexts across stages. This process is similar to traditional practices in creative writing, where creation typically requires multiple rounds of planning, adjustment, and review to gradually perfect the final work \cite{gero2023social}. FIERO's card system offers structured flexibility, meaning it can support the free flow of idea generation while also allowing the use of AI-generated suggestions within the system to help the group focus. For example, group members use the cards to generate characters, scenes, and plots, while AI helps them integrate these fragmented ideas into a coherent story, ensuring logical consistency and providing structured feedback. This combination makes the creative process both free-flowing and directional, potentially meeting diverse creative needs and balancing structure with openness \cite{Chung2022TaleBrush}. The iterative design of FIERO further validates the cognitive models of creativity, which suggest that the iterative process promotes the transformation from initial conceptualization to final structuring \cite{li2018motivational}. While both the Control group and FIERO experienced iterative idea generation during the creative process, FIERO enables exploration of more possibilities during free idea generation, as shown in Section \ref{sec:6.3}.

Many studies on creative narrative emphasize that the depth and emotional connection of a story come not only from fictional fantasy worlds but also from real-life emotions and social experiences \cite{is2019social}. By incorporating elements from real life, such as daily life, social issues, and personal emotions, the creative process not only enhances resonance but also provides audiences with new perspectives. This phenomenon also emerged in the writing of participants using FIERO. Rather than referencing film works, participants tended to start from real-life experiences. For instance, they designed superhero characters with abilities like “housework” or “fashion,” making them not just part of the ``superpowers vs. villains'' model, but also incorporating them into modern societal scenarios, highlighting the unique value and emotional experiences of ordinary people in everyday life. This design provided creators with an inspiration-driven creative space, allowing stories and characters to be more grounded in real life, while also sparking participants' reflections on real-world issues. For example, a participant (G5 Player A) revealed that they faced pressure from their boss at work, so they modeled the villain and personal hero as a conflict between a boss and an employee. The design of FIERO illustrates the potential for stories to gain greater real-life relevance when they address issues like personal growth, social responsibility, and emotional support. 

This tendency to construct narratives based on personal experience resonates with the current trend of using games to explore social issues. Numerous game-based projects demonstrate that games can transform abstract topics into participatory, interactive experiences that spark reflection and dialogue, effectively fostering deep conversations around environmental issues \cite{zhang2025can}, climate change \cite{zhou2024eternagram,zhou_eternagram_2024}, and media bias \cite{tang2025breaking}. \added{The expansion into tabletop games also possesses the potential to trigger real-world reflection, as exemplified by Kaona \cite{baker2024enabling}, a tabletop role-playing game that transforms complex real-world traumas and daily confrontations into reflective quests within the gameplay. By weaving personal experiences into the narrative, players are enabled to combine individual emotional catharsis with collective sociological reflection.} FIERO extends this paradigm into the domain of creative writing. \replace{Its distinctiveness lies in}{A distinctive aspect of FIERO is} its shift from knowledge transmission to active meaning-making, highlighting the potential of games to simultaneously facilitate social discourse and personal creative expression. In FIERO, players become meaning-makers through active reflection, constructing their self-narratives through exploration and expression.

\subsection{The Role of Large Language Models in Collaborative Writing}

\replace{The core effectiveness of FIERO}{A key aspect of FIERO's effectiveness} lies in using LLM to enhance human creativity. The AI in FIERO serves as a supportive collaborator, not dominating the creative process. Research supports that LLMs are effective as neutral collaborators in creative tasks \cite{gero2023social, biermann2022tool}. FIERO's LLM handles two functions: synthesizing fragmented inputs (e.g., merging stories into a subplot) and resolving conflicts (e.g., choosing a villain with reasoning based on heroes). Similarly, SKETCHAR \cite{ling2024sketchar} uses LLM-generated images to help designers refine concepts and communicate with illustrators. Despite different outputs, both use LLM as a mediating tool to lower expertise barriers and build consensus. This shows the versatility of LLM in supporting diverse creative collaboration. Likewise, CharacterMeet's use of GPT-4 for consistent character voices reflects the same principle \cite{qin2024charactermeet}. LLM helps address the common challenge of narrative consistency in group creation \cite{chung2025toyteller}. FIERO's LLM is constrained by the game process: it offers suggestions, not judgments, so humans retain creative control. This aligns with ``augment, not replace'' human agency \cite{gero2023social}. LLM may also enhance accessibility: it translates vague card ideas into polished prose, reducing barriers for less experienced members. This mirrors CoAuthor's finding that LLM democratizes collaboration by bridging skill gaps \cite{Lee2022}. 

However, limitations emerged: LLM reproduced stereotypes (e.g., gender-neutral superheroes framed as male). They also sometimes avoided edgy content: a player from group G8 said AI declined to help create a serial killer character. Additionally, LLM sometimes misunderstood ideas, highlighting the need for bias mitigation \cite{Navigli2023Biases} and human oversight, as LLM reflects training biases \cite{Cheng2023Marked}, which may impact the generation of creative content. \added{We also observed that LLM tends to assimilate unique human linguistic styles. For instance, ``a passionate workplace newcomer who courageously confronts a sinister boss at the company'' was sanitized to ``a passionate workplace newcomer.'' This shows LLM's tendency toward safer expressions \cite{huang2026not}. In creative writing, passively accepting AI-generated phrasing may diminish the human touch—personal emotion, narrative intuition, and cultural heterogeneity. Undeniably, AI possesses powerful capabilities for textual expansion and rhetoric, but its feedback is uncertain with varied raw texts. Therefore, how to enhance expressive quality without sacrificing human autonomy remains key for human-AI co-creative symbiosis in HCI \cite{tang2026human}. This may help mitigate uncertainty while preserving distinct human expression amid intelligent automation.

Beyond LLM limitations, AI interventions may disrupt face-to-face interactions, shifting attention from peers to AI and diminishing social presence \cite{yamashita2026shared}. Although FIERO uses user-initiated triggers, eye contact and discussion were briefly interrupted when players looked at the screen awaiting AI results. Consequently, balancing on-screen content with physical social space remains paramount for future tools.

Notably, while AI image biases are typically seen as problems to fix \cite{roby2025storycrafting}, in FIERO they sometimes activate human senses, offering a fresh perspective on AI limitations. For example, an AI-rendered hero with an extra leg (not matching players' imagination) prompted more nuanced plot twists than text-only prompts could. Player A (G1) said this deviation added a novel trait. This aligns with the ``mild place illusion'' that strengthens creative immersion \cite{Goncalves2018Mild}.}

\subsection{Game-Based Collaboration Enhances Narrative Quality}

Through multi-dimensional evaluation, this study found that the FIERO framework demonstrates significant advantages in story generation, particularly in the three dimensions of plot coherence, character fidelity, and language use (net advantages of 43.2–67.9\%). This may be related to FIERO's game mechanisms. Previous research has found that images can help writers establish clearer semantic associations in narrative organization by making temporal/causal relationships between events explicit \cite{alidoost2014effect}. FIERO further embeds these visual materials within game rules involving random card drawing and forced associations. These visualized narrative anchors compel players to develop ideas around the drawn elements, continuously responding to questions such as ``what happens next'' and ``how would the character react,'' thereby naturally avoiding the fragmented narratives and inconsistent character behaviors commonly found in the Control group. The meaning of a story emerges from the narrative occasion co-constructed by storyteller and audience \cite{gubrium1998narrative}. Although both FIERO and control conditions provided such discussion opportunities, FIERO transforms vague creative negotiation into tangible game-like decision-making through rule-based cards, enhancing the logical convergence of collaborative discussions.

The improvement in the development dimension was relatively modest (23.5–54.3\%), which may be related to the limited number of game rounds. \replace{Research indicates}{Existing research suggests} that the depth of world-building is positively correlated with the number of available events \cite{herman2011basic}. Although FIERO generated more nuanced local settings, the limited rounds in a single game session may lead players to prioritize negotiating ``what happens next'' over reflecting on the world's rules. This finding suggests that future frameworks could design dedicated world-building reflection sessions, encouraging players to continuously deepen and connect with earlier settings across multiple iterations.

The divergent evaluations on the creativity and anthropomorphism dimensions reveal the inherent complexity of FIERO stories on these dimensions. This may be because the two groups exhibited different types of creativity. FIERO excels at conceptual creativity, such as the setting where ``VR glasses completely scramble the colors the villain sees''—such ideas often originate from cards (e.g., specific imagery, actions, or object cards) that force narrators to establish connections between seemingly unrelated elements. This phenomenon reflects how stories can produce emergent content beyond designers' expectations when creation is constrained by non-linear rules such as card drawing \cite{ryan2015narrative}. The Control group, in contrast, excelled at plot creativity, such as the twist where ``the protagonist enters a dream to extract information from the gang.'' Both groups followed linear writing processes, during which writers' thinking tends to exhibit goal-oriented characteristics \cite{kapoor2019creativity}. However, FIERO's card mechanism introduced more non-linear possibilities, facilitating serendipitous associations and thereby generating more conceptually creative outputs.

Despite these dimensional divergences, the three judges reached a high degree of consensus on overall preference (45.7–65.4\% net advantage). FIERO's success suggests that the value of gamified collaboration may stimulate those core qualities that make a story compelling. In other words, FIERO may not produce perfect stories, but it produces stories that are more engaging. For creative writing, this may be the more important criterion.

\subsection{Design Implications}

\subsubsection{Lightweight Scaffolding and Creative Anchors}

The design of FIERO offers several implications for collaborative creative tools. By integrating tangible cards as low-friction idea anchors with digital systems, the game demonstrates how hybrid interfaces can address common challenges in group storytelling. These challenges include fragmented contributions and inefficient decision-making, all of which the system seeks to address without disrupting human interaction. This ``lightweight scaffolding'' provides \replace{necessary}{useful} structural guidance while preserving creative freedom, suggesting that collaborative tools should prioritize concrete, shared stimuli to reduce ambiguity and foster collective ownership of ideas.

\subsubsection{Multimodal Design Enabling Diverse Scenarios}

Furthermore, the game's multimodal design integrates physical manipulation, digital visual representations, and face-to-face discussion. This integration creates crosssensory resonance that helps clarify abstract concepts and deepens immersion. By positioning intelligent systems as ``facilitators'' rather than dominators, the approach allows technology to handle routine integration tasks while leaving creative decisions to humans. This balanced approach amplifies rather than replaces human agency. The framework can be applied to educational settings such as classroom interactions and icebreaker activities, as well as professional co-creation such as themed brainstorming sessions. It also offers potential for crosscultural or mixedlanguage team collaboration \cite{Chen2025Collaborative}.

\subsection{Limitations}

\added{\subsubsection{The Bundled Comparison}

A limitation of this study is the ``bundled'' nature of our comparison: we evaluated FIERO holistically rather than isolating individual design components. We compared FIERO against a Control group using plain collaborative writing in Google Docs. FIERO includes physical cards, AI imagery, decision support, and a tangible-digital interface, while the baseline is just a simple text editor. Although Google Docs reflects real-world practices of non-professional creators, FIERO introduces multiple distinct affordances, adding confounding variables. Consequently, it is unclear whether positive effects (e.g., reduced cognitive load, better sense-making, increased playfulness) come from card mechanics, AI capabilities, or the game framework. This holistic measurement also hides cognitive costs of switching between physical cards and digital UI—cross-modal friction is masked in aggregate scores. We acknowledge this limits the granularity and precision of our findings. Future work could use an unbundled, multi-arm design (e.g., cards-only, AI-only, full FIERO) to isolate each component's value. This would also clarify which modality combinations distribute cognitive resources to reduce mental load, and which compete for the same channels, increasing switching costs.

\subsubsection{Effect Sizes}
The observed effect sizes ranged from small to medium ($|r| = 0.005$ to $0.444$). Although the sample size ($n = 30$ per group) is reasonable for detecting moderate effects, it limits precision for smaller effects. Future research could improve precision in two ways: larger samples would provide more precise estimates, and, as noted in the bundled comparison limitation above, unbundled designs that isolate individual intervention components may yield more precise effect estimates.


\subsubsection{Multiple Comparisons and Type I Error}
As noted in the data analysis section, multiple comparisons at the item level inflate the risk of Type I errors. Because most nominally significant results did not survive the strict Bonferroni correction, these uncorrected findings must be treated with caution as purely exploratory. Crucially, rather than being viewed in isolation, these quantitative tendencies should be interpreted comprehensively alongside the qualitative findings for a more nuanced understanding.}

\subsubsection{Opacity of AI Conflict Resolution}

A limitation of this study is the opacity in how FIERO's AI chooses among conflicting player drafts. Although each selection includes a textual rationale, the criteria and reasoning are not fully transparent in three ways. First, the selection standard is not explicitly defined. The system prompt is told to select the ``most appropriate'' option, but \emph{appropriateness} is interpreted by GPT-4, not a player-defined rubric. Players cannot anticipate which dimensions (e.g., dramatic contrast, coherence, novelty, fit with cards) the AI is weighing. Second, the rationale is a post-hoc justification, not a faithful trace of the selection process. Because the AI generates fluent text based on the chosen output, a confident rationale does not prove those factors drove the decision. Players cannot distinguish a real reason from a plausible rationalization based on the text alone. Third, the system does not surface the AI's uncertainty. Decoding uses temperature 0.7, the same drafts could yield a different selection; players are not shown how close rejected drafts were, nor given a second-choice option. This means the suggestion can appear more decisive than it warrants.

\subsubsection{Game Duration and Narrative Depth}

The card-based mechanics and content generation tools employed in this study help teams quickly construct a story framework within a limited time, lowering the barrier to creation. Within a 30-minute game session, this design effectively supports the generation of complete stories. However, for teams wishing to delve deeper into character development and plot intricacies, time becomes a constraint, making them more reliant on the tools to complete their creation. Future iterations could consider introducing more flexible pacing controls, allowing teams to find a better balance between narrative depth and creative efficiency that suits their needs.

\subsubsection{Participant Background and Narrative Genres}

The participants in this study were primarily university students, and the story creation tasks were centered on the superhero genre. This choice facilitated the testing of the collaborative process within a relatively unified context, but it also limits the generalizability of the findings to other genres (e.g., realistic fiction, fantasy literature) and populations (e.g., professional writers, cross-generational teams). Future research could expand the sample scope to explore collaborative performance under diverse narrative preferences and creative habits. Furthermore, the current evaluation method relies mainly on subjective feedback, which, while capable of reflecting participant experiences, still has room for improvement in measuring creative growth.

\subsubsection{Team Size and Group Dynamics}

This study fixed the team size at three members per group, a choice that enabled observable team interactions and ensured operational feasibility in data collection. However, in real-world collaborative writing scenarios, team sizes are often more flexible—dyads or teams of four or more may exhibit different collaboration patterns and group dynamics. While the triadic structure offers relative stability, it may also obscure phenomena unique to other configurations, such as the deep dialogue characteristic of pair collaboration or the diversity of ideas that can emerge in larger teams. Future studies could investigate how varying team sizes influence both the collaborative process and the quality of co-written narratives.



\subsubsection{Transparency of Content Evaluation Tools}

In the content analysis, we used three different LLM models (Claude, Gemini, and GPT) to evaluate the content, all with the same prompt. However, the internal mechanisms of these models are not fully transparent, meaning we have limited insight into their parameters and training data. While using models with divergent architectures helps mitigate model-specific biases, the possibility of systematic preference for AI-assisted content cannot be fully ruled out \added{\cite{laurito2025ai}}. \replace{Nevertheless, the models provided useful results, and this limitation points to future directions: improved model transparency or more diverse evaluation approaches could further enhance the rigor and fairness of the scoring process.}{Human-in-the-loop evaluation is a necessary mechanism for this type of evaluation. In our current evaluation process, human experts are involved at the initial stage by defining the evaluation criteria that guide the LLM-based assessment. However, humans do not directly control or influence the final evaluation outcomes. Moving forward, rather than relying on either a fully automated workflow or a purely human-led evaluation process, we argue for a human–AI collaborative evaluation workflow for future direction \cite{wen2026ai}. In such a workflow, human judges could assess comparative results with AI support. For each comparison, the AI could generate summaries of the two stories, visual representations of their narrative structures, and detailed scoring rationales for human review. Human judges could then choose to accept the AI-generated evaluation, revise the results, or provide additional comments and instructions for the AI to reconsider the comparison. This approach would preserve human oversight while using AI to reduce evaluation workload, improve consistency, and support more transparent evaluation.}

\section{Conclusion}\label{sec:Conclusion}
In conclusion, this study presents FIERO, a collaborative story creation game that integrates physical card mechanics with a generative AI-powered digital system to support collaborative creative writing. By combining tangible interaction with intelligent assistance, the tool fosters an immersive and visually evocative writing environment that enhances team resonance and \replace{collective meaning making}{the quality of collaborative writing}. It contributes to narrative coherence while helping integrate diverse ideas and supports collective decision-making, enriching both the storytelling process and the depth of group expression. \added{Our exploration demonstrates that embedding real-life emotional and social experiences within a gamified framework empowers creators to anchor imaginative worlds in reflective, real-world commentary. Crucially, while LLMs can function as neutral collaborators within team dynamics, they introduce potential risks regarding text homogenization, stereotypical biases, and the overshadowing of unique human linguistic voices. Ultimately,} FIERO offers a distinct alternative to traditional writing methods, demonstrating the potential of hybrid systems to transform how teams create together.

\bibliographystyle{ACM-Reference-Format}
\bibliography{references}


\clearpage
\appendix

\label{sec:Appendix}
\section{Survey Questionnaire} \label{app:Questionnaire}

\begin{table}[ht]
\caption{Statistical Results for Creativity Support, Creative Self-Efficacy, and Story Satisfaction Assessments (n=30 per group)}
\resizebox{\textwidth}{!}{
\begin{tabular}{@{}l p{3cm} p{3.3cm} l l l@{}}
\toprule
\textbf{Variable} & \textbf{GAME M$\pm$SD} & \textbf{CONTROL M$\pm$SD} & \textbf{z} & \textbf{p-value} & \textbf{Effect size (r)} \\
\midrule
\textbf{Creativity Support Index} & 6.12$\pm$0.53 & 5.79$\pm$0.60 & -2.141 & 0.032* & -0.276 \\
\midrule
\textbf{Creative Self-Efficacy Scale} & 5.31$\pm$0.83 & 4.94$\pm$0.64 & -1.984 & 0.047* & -0.256 \\
\midrule
\textbf{Story Satisfaction Assessment} & 6.01$\pm$0.50 & 5.16$\pm$0.90 & -3.441 & 0.001** & -0.444 \\
\bottomrule
\end{tabular}}

\footnotesize{\textit{Note.} *p<0.05, **p<0.01. z = Mann-Whitney U test statistic; r = z/$\sqrt{N}$ (N=60). Negative r indicates higher scores in GAME group.}
\end{table}

{\small

\begin{longtable}{@{} p{0.1\textwidth} p{0.36\textwidth} p{0.1\textwidth} p{0.1\textwidth} p{0.06\textwidth} p{0.06\textwidth} p{0.08\textwidth} @{}} 
\caption{Statistical Results for Each Item in Creativity Support, Creative Self-Efficacy, and Story Satisfaction Assessments (n=30 per group)} \\
\toprule
\textbf{Variable} & \textbf{Item} & \textbf{GAME M$\pm$SD} & \textbf{CONTROL M$\pm$SD} & \textbf{z} & \textbf{p-value} & \textbf{Effect size (r)} \\
\midrule
\endfirsthead

\toprule
\textbf{Variable} & \textbf{Item} & \textbf{GAME M$\pm$SD} & \textbf{CONTROL M$\pm$SD} & \textbf{z} & \textbf{p-value} & \textbf{Effect size (r)} \\
\midrule
\endhead

\bottomrule
\multicolumn{7}{p{13cm}}
{\footnotesize\textit{Note.} *p<0.05, **p<0.01. z = Mann-Whitney U test statistic; r = z/$\sqrt{N}$ (N=60). Negative r indicates higher scores in GAME group. Note that under Bonferroni correction for 15 exploratory item-level comparisons, the adjusted significance threshold is $\alpha_{new} = 0.0033$; items with $0.0033 < p < 0.05$ indicate non-significant exploratory trends.} \\
\endlastfoot

\multirow{5}{=}{\textbf{Creativity Support Index}}  
& This way of writing kept me engaged in the creative process. & 6.15$\pm$0.95 & 6.19$\pm$0.79 & -0.037 & 0.970 & -0.005 \\
& This writing experience sparked my imagination. & 6.33$\pm$0.78 & 6.19$\pm$0.68 & -0.910 & 0.363 & -0.117 \\
& This writing environment allowed me to fully immerse myself in the creation. & 6.07$\pm$1.04 & 5.81$\pm$0.92 & -1.157 & 0.247 & -0.149 \\
& This way of writing helped me make the story vivid. & 6.26$\pm$1.02 & 5.56$\pm$1.12 & -2.552 & 0.011* & -0.329 \\
& The use of the writing tool felt natural and did not interfere with my creation. & 5.78$\pm$1.40 & 5.19$\pm$1.24 & -1.992 & 0.046* & -0.257 \\
\midrule
\multirow{5}{=}{\textbf{Creative Self-Efficacy Scale}} 
& I believe I can quickly generate a large number of writing ideas. & 5.22$\pm$1.12 & 4.56$\pm$1.15 & -2.171 & 0.030* & -0.280 \\
& I am skilled at building upon others' ideas to develop my own. & 5.81$\pm$1.33 & 5.52$\pm$1.22 & -1.144 & 0.253 & -0.148 \\
& I consider myself to be good at generating novel ideas. & 5.56$\pm$1.09 & 4.93$\pm$1.11 & -2.032 & 0.042* & -0.262 \\
& I am confident that I can produce original and unique work. & 5.59$\pm$1.22 & 5.04$\pm$1.19 & -1.621 & 0.105 & -0.209 \\
& If my teammates do not accept my innovative ideas, I will find ways to persuade them. & 4.37$\pm$1.52 & 4.67$\pm$1.24 & -0.788 & 0.431 & -0.102 \\
\midrule

\pagebreak
\multirow{5}{=}{\textbf{Story Satisfaction Assessment}} 
& The story's plot development was logical and coherent. & 5.96$\pm$0.94 & 5.04$\pm$1.43 & -2.583 & 0.010* & -0.333 \\
& The story's plot and setting were creative and defied my expectations of traditional stories. & 6.04$\pm$1.02 & 5.00$\pm$1.30 & -3.006 & 0.003** & -0.388 \\
& I believe teamwork positively influenced the quality of the story. & 6.33$\pm$0.96 & 5.78$\pm$1.28 & -1.787 & 0.074 & -0.231 \\
& The story I wrote connected well with my teammates' contributions. & 5.93$\pm$1.33 & 5.33$\pm$1.11 & -2.183 & 0.029* & -0.282 \\
& I found it easy to make decisions that benefited the team's story during creation. & 5.81$\pm$1.08 & 4.63$\pm$1.76 & -2.487 & 0.013* & -0.321 \\

\end{longtable}}

\section{Interview Question List} \label{app:Interview}

We conducted semi-structured interviews with all participants. The interviewers followed a common question list (see examples below), which included questions common to both groups as well as specific questions for the experimental or control group. To gain a deeper understanding, the interviewers also asked follow-up questions based on participants' responses.

\subsection{About the Creative Process}
\begin{enumerate}
\item Did you experience any moments of creative block during writing? If so, what helped you move forward?
\item What about your teammates? Did they help inspire new ideas, or did their input sometimes constrain your thinking?
\item Where did you draw inspiration for your character designs and abilities?
\item Were there any specific experiences, stories, or media that influenced your creative decisions?
\item When developing weapons and villains, did you consider broader implications such as social, technological, or ethical dimensions?
\end{enumerate}
\subsection{About Personal Experience}
\begin{enumerate}
\item Did any disagreements arise during your collaboration? If so, how were they resolved?
\item How did you and your teammates organize the writing process?
\item How satisfied are you with the final story your team created?
\item Did you ever feel that your ideas were overlooked during discussions, or that your contributions were not fully valued?
\item Do you feel that your vision for the characters and story was fully realized in the final output?
\end{enumerate}

\section{\added{Demographic Information of Participants}} \label{app:Participants}

A total of 60 participants took part in this study, divided equally into a Control group (n = 30) and a Game group (n = 30). All participants were between 18 and 28 years of age; we collected only the age range, not individual ages. The detailed demographic information, including group assignment, participant ID, gender, academic major and current role, is presented below.

\begin{table}[H]
\centering
\caption{Demographic Details of Control Group Participants}

\begin{tabularx}{\textwidth}{>{\hsize=0.875\hsize}X >{\hsize=0.875\hsize}X >{\hsize=0.875\hsize}X >{\hsize=1.5\hsize}X >{\hsize=0.875\hsize}X}
\toprule
\textbf{Group} & \textbf{ID} & \textbf{Gender} & \textbf{Major} & \textbf{Role} \\
\midrule
1 & C1  & Male   & Product Design                 & Student \\
  & C2  & Female & Product Design                 & Student \\
  & C3  & Male   & Business Administration & Student \\
2 & C4  & Female & Product Design                 & Student \\
  & C5  & Female & Business Administration & Student \\
  & C6  & Male   & Business Administration & Student \\
3 & C7  & Female & Fashion        & Student \\
  & C8  & Female & Industrial Design     & Student \\
  & C9  & Female & Fashion        & Student \\
4 & C10 & Female & Design Studies        & Student \\
  & C11 & Female & Digital Media Technology & Student \\
  & C12 & Male   & Industrial Design     & Student \\
5 & C13 & Female & -                     & Teacher \\
  & C14 & Male   & -                     & Designer \\
  & C15 & Male   & -                     & Software Engineer \\
6 & C16 & Male   & -                     & Video Editor \\
  & C17 & Male   & -                     & Planner \\
  & C18 & Male   & -                     & Planner \\
7 & C19 & Female & Industrial Design     & Student \\
  & C20 & Male   & Design Studies        & Student \\
  & C21 & Female & Digital Media Technology & Student \\
8 & C22 & Female & Fashion        & Student \\
  & C23 & Male   & Architecture          & Student \\
  & C24 & Male   & Architecture          & Student \\
9 & C25 & Female & Industrial Design     & Student \\
  & C26 & Female & Design Studies        & Student \\
  & C27 & Female & Digital Media Technology & Student \\
10 & C28 & Female & Digital Media Technology & Student \\
   & C29 & Male   & Architecture          & Student \\
   & C30 & Female & Design Studies        & Student \\
\bottomrule
\end{tabularx}
\end{table}

\begin{table}[H]
\centering
\caption{Demographic Details of Game Group Participants}
\begin{tabularx}{\textwidth}{>{\hsize=0.875\hsize}X >{\hsize=0.875\hsize}X >{\hsize=0.875\hsize}X >{\hsize=1.5\hsize}X >{\hsize=0.875\hsize}X}
\toprule
\textbf{Group} & \textbf{ID} & \textbf{Gender} & \textbf{Major} & \textbf{Role} \\
\midrule
1 & G1  & Male   & -                        & Research Assistant \\
  & G2  & Female & Computer Science & Student \\
  & G3  & Male   & Computer Science         & Student \\
2 & G4  & Male   & Industrial Design        & Student \\
  & G5  & Male   & Industrial Design        & Student \\
  & G6  & Male   & Industrial Design        & Student \\
3 & G7  & Male   & -                        & Programmer \\
  & G8  & Male   & -                        & Officer \\
  & G9  & Male   & -                        & Officer \\
4 & G10 & Female & Industrial Design        & Student \\
  & G11 & Male   & Design Studies           & Student \\
  & G12 & Female & Architecture             & Student \\
5 & G13 & Male   & -                        & Video Editor \\
  & G14 & Male   & -                        & Sales \\
  & G15 & Male   & -                        & Video Editor \\
6 & G16 & Female & Industrial Design        & Student \\
  & G17 & Female & Digital Media Technology & Student \\
  & G18 & Female & Industrial Design        & Student \\
7 & G19 & Female & Digital Media Technology & Student \\
  & G20 & Female & Fashion           & Student \\
  & G21 & Female & Fashion           & Student \\
8 & G22 & Female & Digital Media Technology & Student \\
  & G23 & Female & Digital Media Technology & Student \\
  & G24 & Female & Industrial Design        & Student \\
9 & G25 & Female & Finance                  & Student \\
  & G26 & Female & Finance                  & Student \\
  & G27 & Female & Education                & Student \\
10 & G28 & Female & Finance                  & Student \\
   & G29 & Male   & Computer Science         & Student \\
   & G30 & Female & Computer Science         & Student \\
\bottomrule
\end{tabularx}
\end{table}

\section{Content Analysis Prompt} \label{app:content}
\noindent\rule{\linewidth}{0.2pt}
\begin{lstlisting}[breaklines=true, basicstyle=\ttfamily]
# Role
You are an expert evaluator of creative writings and stories.
# Task
Compare the following two creative writing samples [Sample A] and [Sample B]. Judge based on story quality, not length.
**Important: Only output concise judgment results (Win/Lose), do not output lengthy analysis.**
# Dimensions
Evaluate the stories based on the following dimensions. For each dimension, clearly understand what is being evaluated, what counts as good (Win), and what counts as poor (Lose):
1. **Plot**: Evaluate the quality and coherence of the story's plot structure.
   - **What to evaluate**: Story arc completeness, narrative flow, plot development logic, conflict setup and resolution, pacing, plot coherence
   - **Win (Good)**: Clear beginning-middle-end structure, logical plot progression, well-developed conflicts with satisfying resolution, smooth narrative flow, coherent storyline
   - **Lose (Poor)**: Confusing or incomplete plot, illogical plot progression, unresolved conflicts, abrupt or unclear transitions, incoherent storyline
2. **Development**: Evaluate the development and setting of the story.
   - **What to evaluate**: World-building depth, character development, setting details and richness, story progression, background information
   - **Win (Good)**: Rich and detailed world-building, well-developed characters with clear backgrounds, vivid setting descriptions, gradual and natural story progression, sufficient background information
   - **Lose (Poor)**: Shallow world-building, underdeveloped characters, vague or missing setting details, rushed or unclear story progression, lack of background information
3. **Language Use**: Evaluate the quality of language and writing style.
   - **What to evaluate**: Writing style sophistication, vocabulary richness, sentence structure variety, clarity of expression, expressiveness, readability
   - **Win (Good)**: Sophisticated writing style, rich and appropriate vocabulary, varied sentence structures, clear and expressive language, engaging and readable prose
   - **Lose (Poor)**: Simple or repetitive writing style, limited vocabulary, monotonous sentence structures, unclear or confusing expression, dull or hard-to-read prose
4. **Anthropomorphism**: Evaluate how well characters are portrayed as human-like.
   - **What to evaluate**: Character depth and complexity, human-like qualities (emotions, thoughts, motivations), relatability, emotional expression, psychological realism
   - **Win (Good)**: Deep and complex characters, rich human-like qualities (emotions, thoughts, motivations), highly relatable characters, vivid emotional expression, psychologically realistic portrayal
   - **Lose (Poor)**: Shallow or one-dimensional characters, lack of human-like qualities, unrelatable characters, weak or missing emotional expression, unrealistic or mechanical portrayal
5. **Character Fidelity**: Evaluate the consistency and fidelity of character portrayal.
   - **What to evaluate**: Character consistency throughout the story, staying true to established character traits, believable behavior, character coherence, no contradictions
   - **Win (Good)**: Consistent character portrayal, characters stay true to their established traits, believable and coherent behavior, no contradictions in character actions or personality
   - **Lose (Poor)**: Inconsistent character portrayal, characters act out of character, unbelievable or incoherent behavior, contradictions in character actions or personality
6. **Overall**: Evaluate overall preference between the two stories.
   - **What to evaluate**: Overall story quality, reader engagement, enjoyment factor, completeness, coherence across all aspects
   - **Win (Good)**: Higher overall quality, more engaging, more enjoyable to read, more complete story, better coherence across all aspects
   - **Lose (Poor)**: Lower overall quality, less engaging, less enjoyable to read, incomplete story, poor coherence across aspects
7. **Creativity**: Evaluate the creativity and originality of the story.
   - **What to evaluate**: Originality of ideas, innovative concepts, creative elements, uniqueness, fresh perspectives, imaginative content
   - **Win (Good)**: Highly original ideas, innovative concepts, creative and unique elements, fresh perspectives, highly imaginative content, breaks new ground
   - **Lose (Poor)**: Common or cliched ideas, lack of innovation, uncreative or conventional elements, unoriginal perspectives, unimaginative content, follows familiar patterns
# Sample A
{sample_a}
# Sample B
{sample_b}
# Output Format (JSON)
**CRITICAL: Only output concise judgment results (Win/Lose), do not output lengthy analysis!**
For each dimension, judge Win (Sample A wins), or Lose (Sample B wins). Reason field must be extremely brief (max 5 characters), only write the core judgment word.
Please strictly follow the following JSON format, do not add any other text:
{
  "Plot": {"Verdict": "Win", "Reason": "More complete"},
  "Development": {"Verdict": "Win", "Reason": "Deeper"},
  "Language Use": {"Verdict": "Lose", "Reason": "Plain"},
  "Anthropomorphism": {"Verdict": "Win", "Reason": "Vivid"},
  "Character Fidelity": {"Verdict": "Win", "Reason": "Consistent"},
  "Overall": {"Verdict": "Win", "Reason": "Better"},
  "Creativity": {"Verdict": "Win", "Reason": "Novel"},
  "Final_Winner": "Sample A"
}
**Important Requirements (Must Strictly Follow):**
- Verdict can only be "Win", or "Lose"
- Final_Winner can only be "Sample A" or "Sample B"
- Reason field must be extremely brief, each max 5 characters, only write the core judgment word (e.g., "More complete", "Plain", "Vivid", "Better", "Novel", etc.)
- **Prohibit lengthy analysis, only output concise judgment results**
- **Do not add any explanatory text, only output JSON format**
\end{lstlisting}
\noindent\rule{\linewidth}{0.2pt}

\section{Group Written prompts}
\label{appendix:Written prompts}
\subsection{Control Group}
Hi everyone, welcome to our collaborative writing test. This test is designed for a creativity-written experiment. Today, we will be using Google Docs to create the story. We will be creating a complete superhero story together, and each of you will contribute to the story with a character you create.

\textbf{Ground 1:} First, each participant will create a background story for one superhero. You are free to create any character—no limits on their powers, appearance, or any other details. You have 5 minutes to complete this task, so be creative! \textit{(5 minutes have passed)} Time's up! Now, please discuss your superhero characters and their abilities with each other. Get to know each other’s creations. \textit{(5 minutes have passed)}

\textbf{Ground 2:} Time's up! Now we move on to the second round. Each participant will think of a location that includes a supervillain. Please create a story involving that location and villain, and you have 5 minutes to complete the task. \textit{(5 minutes have passed)} Time's up! Now, please discuss your created locations and supervillains with each other and choose one story to be included as part of the full story. \textit{(5 minutes have passed)}

\textbf{Ground 3:} Time's up! So, which location did you choose? Now, let's move on to the third round. Each participant will think of an opportunity involving a weapon used by a superhero. Create a detailed story about it, and you have 5 minutes to complete the task. \textit{(5 minutes have passed)} Time's up! Now, please discuss your created opportunities and weapons with each other and choose one story to be included as part of the full story. \textit{(5 minutes have passed)}

\textbf{Final Stage:} Time's up! Which opportunity did you choose? Great job, everyone! Now that we’ve selected the best parts, let's combine them and turn them into a complete story. Thank you all for your participation!

\subsection{Game Group}

Hi everyone, welcome to our card game test. This test is designed for a creativity-written experiment. Today, we will be using FIERO to create the story. We will be creating a complete superhero story together, and each of you will contribute to the story with a character you create.

\textbf{Ground 1:} First, each participant will create a background story for one superhero. Please draw three cards as the characters, according to the elements on the card, you are free to create any character—no limits on their powers, appearance, or any other details. You have 5 minutes to complete this task, so be creative! \textit{(5 minutes have passed)} Time's up! Now, please upload these three cards image in the application, and the system will generate a superhero image based on your description and discuss your superhero characters and their abilities with each other. Get to know each other’s creations. \textit{(5 minutes have passed)}

\textbf{Ground 2:} Time's up! Now we move on to the second round. Each player selects a location card and set as the location of the supervillain. Please create a story involving that location and villain, and you have 5 minutes to complete the task. \textit{(5 minutes have passed)} Time's up! Now, you can upload the card image in the application and discuss your created locations and supervillains with each other and choose one story to be included as part of the full story, or you can click the button to see the suggestions provided by the system (optional). \textit{(5 minutes have passed)}

\textbf{Ground 3:} Time's up! So, which location did you choose? Now, let's move on to the third round. Each participant choose an opportunity card which defined as the weapon used by the superhero. Create a detailed story about it, and you have 5 minutes to complete the task. \textit{(5 minutes have passed)} Time's up! Now, please please upload the card image in the application and discuss your created opportunities and weapons with each other and choose one story to be included as part of the full story or you can click the button to see the suggestions provided by the system (optional). \textit{(5 minutes have passed)}

\textbf{Final Stage:} Time's up! Which opportunity did you choose? Good job, everyone! Now that we’ve selected the best parts, let's combine them and turn them into a complete story. By the way, you can also view the full story provided by the system (optional). Thank you all for your participation!

\section{Prompt Engineering}
\label{appendix:prompts}

Prompt engineering serves as the core mechanism for implementing the system's intelligent assistance. Structured prompt templates were designed to ensure the model outputs align closely with the game's narrative logic. The templates are categorized into three primary types, each corresponding to the distinct collaborative objectives of the three game rounds.

\subsection{\added{Models and Decoding Parameters}}

\begin{table}[h]
\caption{Models and decoding parameters used in FIERO. In-game support models (GPT-4, GPT-4o-image) are deliberately disjoint from post-hoc evaluator models to mitigate same-family preference bias. All in-game calls use a 240\,s request timeout and a 3-attempt connect-retry policy.}
\small
\centering
\begin{tabular}{p{3.8cm}p{2.8cm}p{2.8cm}p{2.0cm}}
\toprule
\textbf{Function} & \textbf{Model} & \textbf{Endpoint} & \textbf{Params} \\
\midrule
Hero / Villain / Weapon image generation & \texttt{gpt-4o-image} & \texttt{/v1/chat/completions} & $T{=}0.7$, ratio $1{:}1$ \\
Villain / Weapon decision support & \texttt{gpt-4} & \texttt{/v1/chat/completions} & $T{=}0.7$, $\textit{max\_tok}{=}1000$ \\
Final-stage story suggestion & \texttt{gpt-4} & \texttt{/v1/chat/completions} & $T{=}0.7$, $\textit{max\_tok}{=}800$ \\
LLM-as-a-Judge (evaluation) & Claude Opus 4.5, Gemini 3 Pro, GPT-5.2 & provider-native & defaults \\
\bottomrule
\end{tabular}
\end{table}

\subsection{\added{System Prompts (Verbatim)}}

The system prompts deployed in FIERO are reproduced verbatim below. They are written to be language-agnostic: the model is instructed to respond in the same language as the user's input.

\begin{itemize}
    \item \textbf{Hero system prompt} —— \textit{You are a professional story-creation assistant specializing in hero characters. Based on the hero story provided by the user, give professional suggestions and improvements covering character background, ability setup, and personality traits. Respond in the same language as the user's input.}

    \item \textbf{Villain system prompt} —— \textit{You are a professional story-creation assistant. Based on the [Hero Stories] written by all players and the [Villain Stories] written by all players, select the villain you consider most appropriate, and provide corresponding story-writing suggestions---for example, explain why this villain is appropriate and what kind of story this villain could develop with the heroes. Respond in the same language as the user's input.}

    \item \textbf{Weapon system prompt} —— \textit{You are a professional story-creation assistant. Based on the [Hero Stories] written by all players, the AI-selected [Villain Story], and the [Weapon Stories] written by all players, select the weapon you consider most appropriate, and provide corresponding story-writing suggestions---for example, explain why this weapon is appropriate and what kind of story this weapon could develop. Respond in the same language as the user's input.}

    \item \textbf{Story-suggestion (Director) system prompt} (final stage) —— \textit{You are a plot director. Strictly follow the hero/ villain/ weapon story content provided by the user for analysis and integration, and only output concise, executable plot-advancement suggestions. Do not fabricate settings that do not exist. Respond in the same language as the user's input.}
\end{itemize}

These prompts encode the design principle that the LLM is positioned as a \emph{supportive collaborator}: its outputs must be grounded in player-authored stories (``strictly follow$\,\ldots$\,do not fabricate settings that do not exist''), and its role is to \emph{select among player inputs} and \emph{justify the selection}, not to author independent narrative content.

\subsection{\added{User-Prompt Assembly Templates}}

Before each AI call, the server concatenates current room state into a structured payload. The Villain-round template is:

\begin{quote}\small\ttfamily
Room ID: \{room\_id\}\\
Timestamp: \{timestamp\}\\
\#Hero stories: \{N\_hero\}, \#Villain stories: \{N\_villain\}\\
\\
{}%
[Hero Stories]\\
Hero P1: \{story\} \;\;...\\
\\
{}%
[Villain Stories]\\
Villain P1: \{story\} \;\;...\\
\\
\{task instruction --- Villain system prompt\}
\end{quote}

The Weapon-round template additionally includes a \texttt{[Weapon Stories]} block and conditions on the AI-selected villain. This template is the only mechanism by which contradictory inputs reach the LLM: the model is asked to (a)~choose one of the player-authored villains/weapons and (b)~explain why, never to invent a fourth alternative.

\subsection{\added{Image, Decision-Support, and Polishing Prompts}}

\begin{itemize}
    \item \textbf{Image Generation Prompt:} \textit{Please generate an image of \$\{subject\} \$\{description\} based on the elements and visual style of the cards. The image should use an aspect ratio of \$\{ratio\}.}
    \item \textbf{Decision Support Prompt:} In the ``Villain'' and ``Weapon'' rounds, the prompt integrates all narrative content from preceding rounds, directing the model to evaluate and suggest based on overall coherence. For example: \textit{Based on all the [Hero Stories] and [Villain Stories] written by the players, select the most suitable villain. Explain the potential dramatic conflicts with the heroes and provide corresponding story-writing suggestions.}
    \item \textbf{Full-Text Polishing Prompt:} During the final synthesis stage, this prompt guides comprehensive editing while preserving authorial voice and plot integrity: \textit{You are a seasoned narrative editor. Please professionally polish the following collaboratively completed story. Requirements: 1) Absolutely retain all character settings, key plots, and original dialogue; 2) Optimize sentence fluency and logical connections; 3) Unify the narrative style and enhance scene immersion; 4) Correct grammar and punctuation errors.}
\end{itemize}

\section{\added{Adversarial Testing and Conflict-Resolution Strategy}}
\label{appendix:adversarial}

FIERO's intelligent-assistance layer ingests \emph{concurrent, intentionally divergent} player drafts. The central design question is therefore not whether the LLM produces fluent prose but whether it remains (a)~\emph{grounded} in player-authored content, (b)~\emph{non-directive}, and (c)~\emph{robust enough that contradiction does not derail the session}. We address each through the design itself rather than through a separate adversarial benchmark.

\textbf{Three-layer conflict-resolution pipeline.} FIERO's resolution mechanism is \emph{not arbitration but suggestion-routing}, split across three loci:

\begin{enumerate}
    \item \textbf{Mechanical resolution (cards as a shared substrate).} Each round begins from a \emph{shared, randomly drawn} card pool (one Location card in Round~2, one Opportunity card in Round~3, drawn collectively). These shared cards constrain the input space \emph{before} the LLM sees it, so the three drafts share environmental and dramatic anchors. The card layer is the first, lightweight conflict-mediator.
    \item \textbf{AI-side resolution (selection with rationale).} Inside the model call, the system prompt forbids invention and asks the LLM to \emph{select} one of the three player drafts and \emph{justify} the selection by referencing elements of all three. The output is surfaced as a suggestion with rationale visible to all three players.
    \item \textbf{Human-side resolution (face-to-face discussion).} The decisive resolution step is the discussion phase that follows the AI suggestion. Players may accept the AI's choice, override it, blend elements, or ignore the suggestion entirely. The shared storyboard does not lock to the AI's selection.
\end{enumerate}

This three-layer pipeline (\emph{shared cards $\rightarrow$ AI-as-icebreaker $\rightarrow$ human decision}) is what we mean by FIERO resolving conflict: conflict is \emph{funneled, not eliminated}, and the human-decision layer always has the last word. Notably, in every observed FIERO session, the final villain or weapon adopted into the shared storyboard could be traced back to a player-authored draft instead of an AI-only invention, which we read as empirical confirmation that the grounding constraint held under real gameplay.


\section{\added{Reproducibility Artifact and Data Release}}
\label{appendix:artifact}

To support reproducibility and reuse, all FIERO artifacts are made openly accessible in our GitHub repository (\url{https://github.com/zxioke/FAIero-game}). 

\textbf{System artifact.} The full codebase (Node.js + Express + Socket.IO server, native HTML/CSS/JS client, AI-orchestration module \texttt{ai-service.js}, configuration scaffolding) is openly available under the MIT license on GitHub. The repository includes \texttt{server.js}, \texttt{ai-service.js}, \texttt{ai-config.js}, \texttt{script-simple.js}, \texttt{index.html}, \texttt{styles.css}, \texttt{i18n.js}, the launch scripts (\texttt{start.sh} / \texttt{start.bat}), and a \texttt{README.md} describing installation, local-network deployment, and how to swap in an alternative OpenAI-compatible LLM endpoint by overriding \texttt{BASE\_URL} and \texttt{API\_KEY} in \texttt{ai-config.js}.

\textbf{Card files.} High-resolution PNG and SVG files for all 20 physical cards (12 Character, 3 Location, 5 Opportunity) are included under \texttt{data/images/cards/} together with print-ready PDF imposition templates (poker-sized, $63 \times 88$\,mm).

\textbf{Prompt files.} A \texttt{prompts/} directory contains the four in-game system prompts (\texttt{hero.txt}, \texttt{villain.txt}, \texttt{weapon.txt}, \texttt{director.txt}), the user-prompt-assembly templates (\texttt{villain\_user.tmpl}, \texttt{weapon\_user.tmpl}), the image-generation prompt (\texttt{image.tmpl}), the LLM-as-Judge content-analysis prompt, and the participant-facing experimenter scripts for Control and FIERO conditions.


\end{document}